\documentclass[
reprint,
onecolumn,         
notitlepage,       
superscriptaddress,
amsmath,amssymb,
aps,
prl,
floatfix,
nofootinbib,
nobibnotes
]{revtex4-2}

\usepackage{xr-hyper} 
\usepackage[
    colorlinks=true,
    linkcolor=blue,
    citecolor=blue,
    urlcolor=blue
]{hyperref}           

\makeatletter
\newcommand*{\addFileDependency}[1]{
  \typeout{(#1)}
  \@addtofilelist{#1}
  \IfFileExists{#1}{}{\typeout{No file #1.}}
}
\makeatother

\newcommand*{\myexternaldocument}[1]{%
    \externaldocument{#1}%
    \addFileDependency{#1.tex}%
    \addFileDependency{#1.aux}%
}

\myexternaldocument{supplementary}
\setcitestyle{super}
\usepackage{url}
\usepackage{siunitx}
\usepackage[utf8]{inputenc}
\usepackage[T1]{fontenc}
\DeclareUnicodeCharacter{0229}{\k{e}}
\usepackage{graphicx}
\usepackage{dcolumn}
\usepackage{bm}
\usepackage{xcolor}
\usepackage{braket}
\usepackage{subfigure}
\usepackage{anyfontsize}
\usepackage{lineno}
\usepackage[mode=buildnew]{standalone}
\usepackage[maxfloats=256]{morefloats}
\begin{document}

\preprint{}
\title{Absence of a Superradiant Phase Transition in Dirac Landau Polaritons}
\author{Elsa Jöchl}
\affiliation{Institute for Quantum Electronics, ETH Zürich, Zürich 8093, Switzerland}
\affiliation{Quantum Center, ETH Zürich, Zürich 8093, Switzerland}
\author{Felix Helmrich}
\affiliation{Institute for Quantum Electronics, ETH Zürich, Zürich 8093, Switzerland}
\affiliation{Quantum Center, ETH Zürich, Zürich 8093, Switzerland}
\author{Frieder Lindel}
\affiliation{Quantum Center, ETH Zürich, Zürich 8093, Switzerland}
\affiliation{Institute for Theoretical Physics, ETH Zürich, Zürich 8093, Switzerland}
\author{Lucy Hale}
\author{Lorenzo Graziotto}
\affiliation{Institute for Quantum Electronics, ETH Zürich, Zürich 8093, Switzerland}
\affiliation{Quantum Center, ETH Zürich, Zürich 8093, Switzerland}
\author{Mona Jarrahi}
\affiliation{Terahertz Electronics Laboratory, UCLA, Los Angeles 90095, United States}
\author{Tobia F. Nova}
\affiliation{Department of Quantum Matter Physics, University of Geneva, Geneva 1211, Switzerland}
\author{Jérôme Faist}
\author{Giacomo Scalari}
\affiliation{Institute for Quantum Electronics, ETH Zürich, Zürich 8093, Switzerland}
\affiliation{Quantum Center, ETH Zürich, Zürich 8093, Switzerland}

\begin{abstract}
\textbf{ One of the most striking predictions in cavity quantum electrodynamics is the condensation of photons into a macroscopically populated ground state, the so-called superradiant phase transition (SRPT). \cite{hepp_superradiant_1973,wang_phase_1973} SRPTs are theorized to occur in light-matter coupled systems above a critical coupling strength, yet have not been experimentally realized in equilibrium. On the contrary, the very existence of SRPTs has been largely disputed by \textit{No-Go theorems}. \cite{rzazewski_phase_1975,knight_are_1978,gawedzki_no-go_1981}
In cavity-coupled electronic systems with Dirac dispersion, the diamagnetic $\vec{A}^2$-term crucial to \textit{No-go theorems} is not present at leading order, making graphene Landau level transitions ultrastrongly coupled to terahertz cavities good candidates for SRPTs. \cite{hagenmuller_cavity_2012,chirolli_drude_2012} In this work, we present the first terahertz spectroscopic measurements of an hBN-encapsulated monolayer graphene flake coupled to a highly sub-wavelength resonator mode. By tuning the graphene carrier density, we drive the resulting Landau polaritons into the ultrastrong coupling regime, with the normalized coupling reaching $\approx 40 \%$, approaching criticality. In this regime, the continuous SRPT would lead to a unique spectroscopic polariton softening, which we consistently rule out. The full polariton dispersion is instead quantitatively reproduced by a Hopfield Hamiltonian using a quasistatic near-field model that accounts for the sub-wavelength character of the cavity. }
\end{abstract}

\maketitle
\section{Introduction}
\linenumbers
\begin{figure*}
    \centering
    \includegraphics[scale= 0.95]{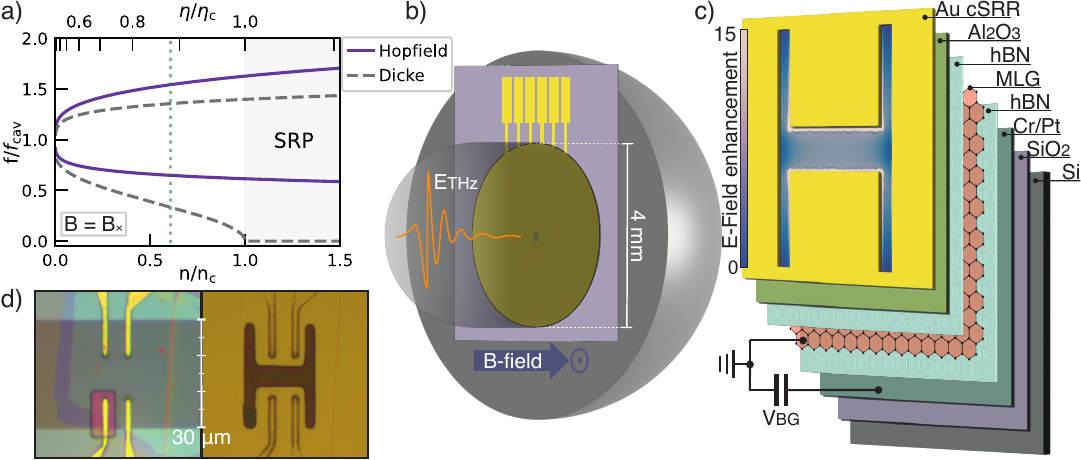}
    \caption{\textbf{Probing SRPTs in monolayer graphene. }
    \textbf{a)} Dispersion of the lower and upper polariton branches as a function of density $n$ (coupling $\eta$) at resonance in the Dicke (grey dashed lines) and Hopfield models (purple lines). While the Dicke prediction exhibits criticality with softening to zero frequency of the lower branch, the Hopfield model does not. The density range where a superradiant phase (SRP) is predicted in the Dicke model is indicated by the grey background, while the maximum density measured in this work is indicated by the green dotted line.  
    \textbf{b)} Schematic view of the experimental setup. The sample consisting of the graphene flake and the cSRR is mounted between Si immersion lenses. A linearly polarized THz pulse $E_\text{THz}$ (orange) is incident on the sample. A static magnetic field (purple) orthogonal to the graphene plane quantizes the electron motion in graphene to Landau levels.  
    \textbf{c)} Precise stack-up of the sample. The graphene flake is encapsulated by hBN flakes and connected electrically using edge contacts. A thin metallic Cr/Pt layer acts as an electrostatic gate by applying a static voltage $V_\text{BG}$ with respect to graphene. The cSRR is fabricated from Au on top of the graphene heterostructure and is electrically insulated from the latter using alumina (Al$_2$O$_3$) dielectric. The distribution of the in-plane electric field obtained by finite-element simulations. The substrate consists of Si with a 285 nm oxide layer. 
    \textbf{d)} Top view of the sample before (left) and after (right) the deposition of the cSRR. }
    \label{fig:schematic}
\end{figure*}

A superradiant phase transition (SRPT) is a second-order quantum phase transition in an electromagnetic cavity strongly coupled to a material system with an optically active transition. \cite{hepp_superradiant_1973,wang_phase_1973} The corresponding ordered superradiant phase (SRP) is characterized by the buildup of a macroscopically populated, coherent photonic ground state above a critical light-matter coupling strength $\eta= \frac{\Omega_R}{\omega_\text{cav}} >\eta_c$, where $\Omega_R$ and $\omega_{\text{cav}}$ respectively denote the Rabi and cavity frequency. \cite{hepp_superradiant_1973, wang_phase_1973, kirton_introduction_2019, schlawin_cavity_2022}

Such SRPTs have been experimentally realized in driven-dissipative systems with external pumping necessary to induce the phase transition. \cite{black_observation_2003, baumann_dicke_2010, zhiqiang_nonequilibrium_2017} Meanwhile, solid-state cavity quantum electrodynamics (cavity QED) is a promising platform to realize a SRPT in equilibrium. \cite{reviewcongsuperrad2016} In this framework, large carrier densities and sub-wavelength cavities with strong electric field confinement naturally enhance the vacuum light-matter coupling to values $\eta > 0.1$. This allows the exploration of ultrastrong light-matter coupling regimes \cite{frisk_kockum_ultrastrong_2019,forn-diaz_ultrastrong_2019}, which are necessary to reach critical coupling without external pumping. Notably, the properties of such systems, particularly cavity-coupled cyclotron transitions, have been studied optically \cite{scalari2012}, and have exhibited anomalous quantum Hall signatures. \cite{enkner_tunable_2025, graziotto_cavity_2026} However, while an equilibrium magnon-spin SRPT-analogue has been reported in Ref. \cite{Kim_Kono_SciAdv_2025}, the experimental realization of a photonic equilibrium SRPT remains an open challenge. 

The theoretical framework leading to the postulation of the SRPT is the Dicke model of cavity QED. \cite{dicke_coherence_1954,hepp_superradiant_1973,wang_phase_1973} The key assumption that allows for this criticality is that the diamagnetic (or $\vec{A}^2$-) term of the interaction Hamiltonian is negligible (with the electromagnetic vector potential $\vec{A}$ describing the cavity field). This would later be disputed by recognizing that the very condition for the SRPT, namely high coupling strength, leads to considerable diamagnetic contributions, resulting in a now broadly established \emph{No-Go theorem}. \cite{rzazewski_phase_1975,knight_are_1978,gawedzki_no-go_1981} Neglecting this term a priori therefore breaks electromagnetic gauge invariance and fundamental sum-rules of quantum mechanics. \cite{rzazewski_phase_1975,knight_are_1978,gawedzki_no-go_1981, nataf_no-go_2010, de2018breakdown, de2018cavity, rokaj_free_2022} At sufficiently high coupling strengths, the $\vec{A}^2$-term is therefore routinely retained, and results in the gauge-invariant Hopfield model \cite{hopfield_theory_1958}, in which the SRPT does not occur. 

Interest in the novel phase transition has nonetheless inspired a vast amount of theoretical predictions and counter-arguments about systems in which the \emph{No-Go theorem} could be circumvented. \cite{nataf_no-go_2010, DeLiberato2013,vukics_elimination_2014,  gulacsi_floquet_2015, jaako_ultrastrong-coupling_2016, andolina_cavity_2019, mazza2019,stokes_uniqueness_2020} The ultrastrong interaction of graphene cyclotron transitions with a terahertz (THz) cavity mode is a prime example of such debate.\cite{hagenmuller_cavity_2012, chirolli_drude_2012} Owing to the linear dispersion of Dirac carriers in graphene, the diamagnetic term does not arise directly in a minimal coupling approach \cite{hagenmuller_cavity_2012}, but is only generated when considering interband transitions. \cite{chirolli_drude_2012}

In the strong-coupling regime of cavity QED, cavity and matter modes hybridize to form quasi-particles known as polaritons. A SRPT is a continuous (second-order) phase transition and is therefore preceded by the softening towards zero-frequency of the lower polariton (LP) branch in the THz spectrum (see Figure \ref{fig:schematic}a) as the carrier density $n$ (and thus the coupling strength $\eta$) is increased. Accounting for diamagnetic contributions, the LP asymptote would, however, stay fixed to the cavity frequency. Thus, the polariton spectrum is a direct signature of the SRPT and provides a clear distinction based on the Dicke and Hopfield-like descriptions.

In this work, we aim to settle the debate of an arising SRPT in graphene Landau polaritons experimentally. While broadband far-field spectroscopy of large-scale graphene flakes is routinely performed in the THz range \cite{valmorra_low-bias_2013, ren_terahertz_2012, ZanottoAPl2015}, these chemically deposited 2D materials usually suffer from poor carrier mobility and are therefore not suitable for studying phenomena such as cyclotron transitions. \cite{de_fazio_high-mobility_2019} On the other hand, the limiting factor in optically probing exfoliated 2D materials with high mobility at THz frequencies is the size of the flakes that can be extracted, which is typically in the range of $\approx$~\SI{10}{\micro\meter}, well below the THz wavelength on the order of ($\lambda \approx 1\,\textrm{mm}$). Studies in the THz range of exfoliated flakes have therefore been restricted to complex near-field techniques such as on-chip spectroscopy, photocurrent measurements, scanning near-field optical microscopy, scanning tunneling microscopy, and experiments using spintronic emitters. \cite{zhao_observation_2023, kipp_cavity_2025,HAN:PhysRevLett2021_photocurr, Ma2023,Cocker2021,vonHoegen2026} 

We perform broadband THz spectroscopy on a gate-tunable graphene heterostructure using our recently demonstrated technique for far-field spectroscopy of sub-wavelength cavities. \cite{rajabali_ultrastrongly_2022, helmrich_cavity-driven_nature_2026,jochl_gate-tunable_2026} This technique relies on the use of an asymmetric solid immersion lens system (see Figure \ref{fig:schematic}b). In this way, we can resolve the ultrastrong coupling of a single complementary split ring resonator (cSRR) to a high-mobility exfoliated graphene flake in the THz regime (see Figure \ref{fig:schematic}c). We present the change in the graphene Landau polariton dispersion over a broad range of carrier densities (see green dotted line in Figure \ref{fig:schematic}a), approaching the critical coupling $\eta_c$. Across the whole range, we show excellent agreement with a Hopfield Hamiltonian derived for the sub-wavelength nature of the cSRR mode, and observe no LP softening. We can therefore rule out the occurrence of a Dicke-SRPT in graphene Landau polaritons.

\begin{figure*}
    \includegraphics[width=\textwidth]{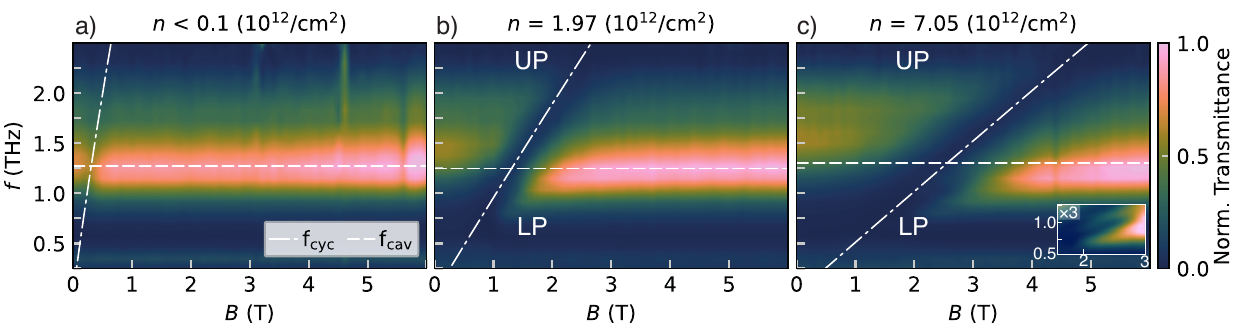}
    \caption{\textbf{Tuning of cyclotron dispersion and light-matter coupling in monolayer graphene.} THz transmission spectroscopy as a function of the magnetic field with monolayer graphene biased at \textbf{a)} the CNP, \textbf{b)} an intermediate density $n =$ \qty{1.97}{10^{12}/cm^{2}}, and \textbf{c)} $n =$ \qty{7.05}{10^{12}/cm^{2}}, performed at a temperature of 3 K. The fitted cavity and cyclotron frequencies are indicated in each spectral map as white dashed and dashed-dotted lines, respectively. All data are referenced to the transmission spectrum of the bare SiO$_2$/Si substrate. The contrast of the color map is normalized to its maximum in each map separately. The inset of panel c) highlights a splitting of the LP branch by enhancing the contrast by a factor of 3.}
    \label{fig:cmaps}
\end{figure*}

\section{Sample and Measurement Technique}
We fabricate a sample consisting of a high-quality graphene van der Waals heterostructure embedded in a gold cSRR resonant at $f_{\text{cav}}\approx \SI{  1.3}{THz}$. The heterostructure and resonator are separated by a distance of $\approx\SI{100}{nm}\ll\lambda$, allowing for near-field coupling. The heterostructure comprises a fully encapsulated hBN - monolayer graphene - hBN stack, placed atop a thin chromium/platinum back gate (see Figure \ref{fig:schematic}c). The gate material is chosen to be transmissive at THz frequencies while being sufficiently conductive for electrostatic gating. The sample is equipped with three edge contacts \cite{wang_one-dimensional_2013} to monolayer graphene (Figure \ref{fig:schematic}d, left), with which we characterize the sample electrically. The gold THz resonator is fabricated on top of the heterostructure (see Figure \ref{fig:schematic}d, right). It is electrically insulated from the graphene heterostructure by a layer of dielectric alumina grown by atomic layer deposition. The field profile of the cSRR is calculated using the finite-element method and is depicted in Figure \ref{fig:schematic}c. It is concentrated in the horizontal gap of the cSRR where the field interacts with graphene carriers. Details of the sample fabrication are provided in the Supplementary Information. 

To couple far-field THz radiation onto the coupled graphene-resonator sample, we employ an asymmetric solid-immersion lens (aSIL) technique \cite{rajabali_ultrastrongly_2022, helmrich_cavity-driven_nature_2026, jochl_gate-tunable_2026}. The sample is mounted between two Si solid immersion lenses (see Figure \ref{fig:schematic}b). These lenses focus the radiation sufficiently to excite the resonator mode and couple the transmitted light to the far-field. Importantly, by extending the metal screen to the full clear aperture of the front lens ($d\approx\SI{4}{mm}$, Figure \ref{fig:schematic}b), only THz radiation interacting with the resonator mode and graphene heterostructure is transmitted, which enhances the sensitivity of our method.

For THz spectroscopy experiments, the sample, including immersion lenses, is placed in a cryostat with optical access coupled to a THz time-domain spectrometer in a transmission geometry. The bandwidth of the spectrometer extends from 0.1 to 2.5 THz. We design the resonator so that the resonance frequency lies in the center of this bandwidth to further enhance sensitivity and ensure a well resolved cyclotron resonance.
The base temperature of the cryostat is \SI{2.8}{K}. Using a  superconducting split-coil magnet, we apply magnetic fields of up to 6 T perpendicular to the sample and parallel to the optical axis (see Figure \ref{fig:schematic}b). 

As a first step, we characterize the sample electrically. We sweep the back-gate voltage $V_\text{BG}$, while recording the source-drain current through graphene (see Figure \ref{fig:schematic}c for a schematic of the electric connections). From the current minimum, we then determine the charge neutrality point (CNP) at $V_{\text{BG}}=V_{\text{CNP}}$. By changing the back-gate voltage, we can capacitively inject either electrons or holes into graphene and thus tune the charge carrier density $n$, as described by a plate capacitor model $n = \epsilon_0\epsilon_{\text{hBN}} V /(ed)$, with the effective gate voltage $V = V_{\text{BG}}-V_{\text{CNP}}$, and the hBN thickness $d =$\qty{25}{nm} (see Supplementary Figure S4). This serves as the reference for subsequent THz spectroscopy experiments at finite doping. We also perform transport measurements at finite magnetic field in a three point-contact quantum Hall (QH) geometry (see Supplementary Figure S3). These measurements show clear quantization of the resistance as a function of the density for a fixed magnetic field. This measurement confirms that the sample quality is sufficiently high for Landau quantization. Thus, we can expect to observe cyclotron resonance in a THz transmission experiment.

\section{Results and Discussion}
We perform THz time-domain spectroscopy for fixed graphene carrier densities $n$ as a function of the applied magnetic field $B$. In Figure \ref{fig:cmaps}, we present exemplary spectral maps of measurements in the low (panel a), intermediate (panel b), and high (panel c) carrier density regimes. In these maps, the color contrast corresponds to the transmitted THz power referenced to the spectrum transmitted through the bare substrate. The full set of spectra is given in the Supplementary Figure "2.

At very low density (close to the CNP), shown in Figure \ref{fig:cmaps}a, a narrow and faint absorption line (highlighted by the white dashed-dotted line) crosses the broader cavity transmission peak around the magnetic field $B_{\times}=0.5$ T. This absorption feature corresponds to the cyclotron absorption of graphene. We do not observe an avoided crossing between the two modes at this density due to the low oscillator strength. Beyond the resonant features of interest, the spectrum shows no significant magnetic-field dependence. The sharp vertical features above $B=\qty{3}{T}$, observed here and in Fig.~\ref{fig:cmaps}c, arise from optical source instabilities and carry no physical significance. Similarly, features below \qty{500}{GHz} are independent of both carrier density $n$ and magnetic field $B$ and are therefore excluded from further analysis. 

When the graphene carrier density is increased to $n$ = \qty{1.97E12}{cm^{-2}}, a polariton splitting becomes visible, centered at the resonant field $B_{\times}=$ \qty{1.4}{T}, as shown in Figure \ref{fig:cmaps}b. The upper polariton (UP), visible at magnetic fields below $B=$ \qty{2}{T}, appears blue-shifted in frequency away from the cavity frequency. Moreover, the lower polariton (LP) is visibly bending from as low as \qty{750}{GHz}, up to the cavity frequency at high magnetic field. 

Finally, in Figure \ref{fig:cmaps}c, we present the spectral map at the highest achievable density of $n$ = \qty{7.05E12}{cm^{-2}}. The anti-crossing field has shifted to $B_{\times}=$ \qty{2.6}{T}, the UP is blue shifted even further, and the LP is asymptotically approaching a similar cavity frequency as in the previous cases. Note that the splitting of the two polariton branches is nearly \SI{1}{THz} at resonance, corresponding to a normalized light-matter coupling $\eta=\Omega_R/\omega_\text{cav}$ of $\approx 40\%$, placing our system well into the ultrastrong coupling regime ($\eta>10\%$) \cite{frisk_kockum_ultrastrong_2019,forn-diaz_ultrastrong_2019}. An emergent feature is highlighted in the inset of Figure \ref{fig:cmaps}c with increased contrast. The LP appears to be split into two distinct branches close to $B_{\times}$. This splitting is consistently observed for high graphene carrier densities (Supplementary Figure S9) and could stem from separate plasmonic modes interacting independently with the cavity. Further discussion of the origin of the splitting is beyond the scope of this work.

\begin{figure*}
    \centering
    \hspace{-1cm}
    \includegraphics[scale= 1]{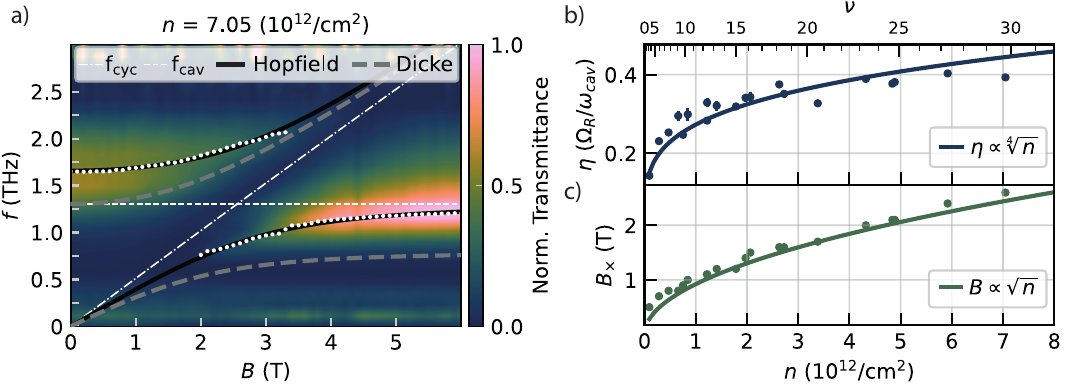}
    \caption{\textbf{Density dependence of ultrastrong light-matter coupling in graphene. } \textbf{a)} Magnetic field-dependent spectral map at an electron density $n =$ \qty{7.05}{10^{12}/cm^{2}} overlaid with the Hopfield model fit (black line) to the fitted peak positions of the respective spectra (white dots). The gray dashed line shows the expected dispersion for UP and LP resulting from a Dicke Hamiltonian at the same cavity frequency and coupling strength. 
    \textbf{b)} Coupling strengths $\eta$ extracted from spectral measurements using the Hopfield-model fitting method as a function of density and filling factor, together with the analytical expression obtained from the quasi-static model . \textbf{c)} Extracted positions of the anti-crossing magnetic field $B_\times$, overlaid with the analytic expression given in Eq. \ref{eq:2}.}
    \label{fig:fig3_fitting}
\end{figure*}

Comparing the three spectral maps, we find that the slope of the cyclotron absorption becomes shallower as the density is increased. Thus, the anti-crossing magnetic field $B_\times$ between the cyclotron and cavity modes increases. This is a clear consequence of the non-parabolicity of the linear band structure of graphene, and stands in stark contrast to conventional semi-conductor materials with a fixed cyclotron dispersion. To explain the dependence of the cyclotron absorption slope on the carrier density, we must first define the cyclotron, or inter-Landau level (LL), transition energy between subsequent LLs. If we assume the filling factor to count spin- and valley degenerate states (see Supplementary Figure S8), they are given by $ \hbar \omega_{\text{cyc}}  = \sqrt{2 v_F^2eB \hbar} \big(\sqrt{\nu+\frac{1}{2}} - \sqrt{\nu -\frac{1}{2}}\big) $. At high filling factor, $\nu \gg 1$, this can be approximated as \cite{Benhamou-Bui_PRL_2026}
\begin{equation}
    \hbar \omega_\text{cyc}  \simeq \sqrt{2 v_{\text{F}}^2eB \hbar} \, \frac{1}{2\sqrt{\nu} }  = \hbar \frac{eB}{m^*} 
    \label{eq:1}.
\end{equation}
Here, we assume a Fermi velocity of $v_{\text{F}} = $ \qty{1E6}{m/s}, and we have introduced the effective cyclotron mass of graphene $m^* = E_{\text{F}} / v_{\text{F}}^2$ that depends linearly on the Fermi energy $E_{\text{F}}$ and thus on $\sqrt{n}$ (see Supplementary Information). The change of $\omega_\mathrm{cyc}$ with $n$ is evidenced by the spectra presented in Figure ~\ref{fig:cmaps}, where the dashed-dotted white lines represent the fitted cyclotron dispersion $\omega_\mathrm{cyc}$. The effective cyclotron mass increases from $m^* \approx 0.01 \times m_e$ in Figure ~\ref{fig:cmaps}a, to $m^* \approx 0.03 \times m_e$ in Figure ~\ref{fig:cmaps}b, to the highest measured value of $m^* \approx 0.05 \times m_e$ in Figure ~\ref{fig:cmaps}c. 

While the cyclotron dispersion changes its slope with the density, the cavity frequency $\omega_{\text{cav}}$ is fixed by the resonator geometry. This causes the magnetic field value at which $\omega_{\text{cyc}}$ anti-crosses with $\omega_{\text{cav}}$ to shift, as the Fermi energy is changed. By rearranging equation \ref{eq:1}, we infer this dependence of magnetic field on the filling factor
\begin{equation}
    B_{\times} = \frac{2 \hbar \omega_{\text{cav}}^2}{e v_F^2} \nu \,.
    \label{eq:2}
\end{equation}
This allows us to solve for the filling factor at resonance $B_\times$, and obtain its resulting square-root dependence on the carrier density $\nu_\times = \sqrt{n \pi v_F^2/g_s \omega_{\text{cav}}^2}$ (see Supplementary Information).

Another general trend we can infer from the spectral maps in Figure \ref{fig:cmaps} is an increase in the polariton splitting with density. The increased splitting leads to a progressively blue-shifted UP, while the LP asymptote remains fixed at the resonator frequency of $\approx\SI{1.3}{THz}$. This behavior is in line with expectations of strongly coupled light-matter systems, where the normalized light-matter coupling strength $\eta$, determining the splitting, is enhanced for larger carrier densities $n$.

In contrast to parabolic semiconductor materials, where the coupling grows as $\eta \sim \sqrt{n}$, the coupling strength in graphene exhibits an unconventional $\sqrt[4]{n}$ dependence on the carrier density. The root cause for this lies in the dependence of the Rabi frequency $\Omega_R = \mu \sqrt{\mathcal{N}}E_{\text{vac}} $ on the electric dipole moment of the coupled transition $\mu \sim 1/(\omega_{\text{cyc}})$ \cite{yao_nonlinear_2013}, as well as the LL degeneracy $\mathcal{N}\sim B$ \cite{hagenmuller_cavity_2012}. As the electric vacuum field $E_{\text{vac}}$ of the cavity stays fixed, we can disregard it here to obtain the dependence of the coupling strength at resonance on the carrier density 
\begin{equation}
    \Omega_R \sim \frac{\sqrt{B_\times}}{\omega_{\text{cyc}}} \sim \frac{m^*}{\sqrt{B_\times}} \sim \sqrt[4]{n}.
\end{equation}
The exact expression derived for Landau polaritons coupled to sub-wavelength cavities can be found in the Supplementary Information, with assumptions of the model detailed later. Importantly, we take into account that the anticrossing magnetic field $B_{\times}$ changes with the filling factor $\nu$ (Eq. \ref{eq:2}), and therefore with $\sqrt{n}$, as detailed above.

\begin{figure*}
    \hspace{-1cm}
    \includegraphics[width= \linewidth]{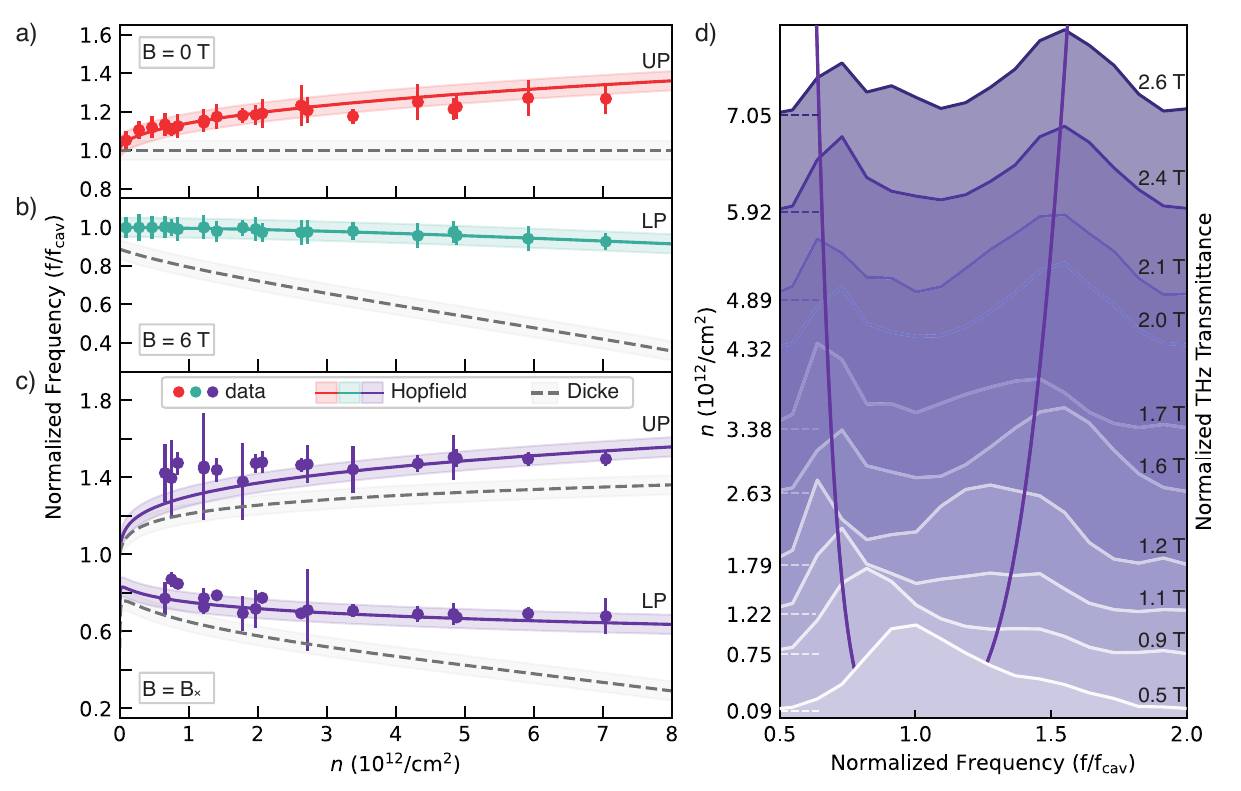}
    \caption{\textbf{Absence of SRPT.} \textbf{a-c)} Normalized peak frequencies extracted from Lorentzian fits for a) the UP asymptote at $B = \qty{0}{T}$, {b)} the LP asymptote at $B = \qty{6}{T}$, and {c)} the UP and LP at the anticrossing magnetic field $B_\times$ for each respective density. Hopfield and Dicke model predictions are overlaid for each case in solid colored and dashed gray lines, respectively. The shaded area indicate the spectral resolution of the measurements. d) Measured THz transmittance spectra at B$_\times$ (varying with density, as indicated on the right). The y-offset corresponds to the density to show the evolution of the UP and LP splitting. Each spectrum is normalized separately. The purple lines correspond to the Hopfield model prediction (also depicted in panel c)).}
    \label{fig:fig4-Hopfield}
\end{figure*}

To analyze our spectra quantitatively, the question arises of whether the polariton dispersions are better fitted by a Hopfield-like model (such as Ref.\,\cite{chirolli_drude_2012}) or by a Dicke-type model as in Ref.\,\cite{hagenmuller_cavity_2012}. In Figure \ref{fig:fig3_fitting}a, we show the spectra at maximum density together with the polariton dispersions for both the Hopfield and Dicke models in solid black lines and dashed gray lines, respectively. The solid black line clearly fits our data well, whereas the dashed gray line does not. This comparison suggests a Hopfield-like dispersion of our system, and we have thus opted to perform a systematic fitting of our data with the Hopfield model. The spectra are first fitted with Lorentzian line shapes to extract the UP and LP peak positions, represented by white dots in Figure \ref{fig:fig3_fitting}a. Then, the best fit is estimated for each measurement by minimizing the root mean square (RMS) error for different cavity frequencies, carrier densities, and coupling strengths. The fitting procedure, along with all optimally fitted dispersions, is displayed in the Supplementary Figure S6. This fitting procedure enables us to report the resulting normalized coupling ratio $\eta$ and anticrossing fields $B_{\times}$ as a function of carrier density (see Figure \ref{fig:fig3_fitting}b-c). We note that we observed no asymmetry in the light-matter coupling between electron and hole doping (see Supplementary Information), and we thus display data points for both electron and hole doping in Figure \ref{fig:fig3_fitting}b-c. The absence of electron-hole asymmetry agrees with expectations, as previous studies on Landau quantized graphene transport predict asymmetries well below our spectral resolution \cite{zhang_experimental_2005}. The densities extracted from the cyclotron slopes are in agreement with the aforementioned plate capacitor model (see Supplementary Figure S7). We can conclude that the expected scaling laws of the coupling strength $\eta \sim \sqrt[4]{n}$ (\ref{fig:fig3_fitting}b) and $B_\times \sim \sqrt{n}$, (exact expression Eq. \ref{eq:2}, overlaid as a solid green line in Figure \ref{fig:fig3_fitting}c) are reproduced by the data extracted from the fits. The exact expression to calculate the coupling strength arises from considerations that will be explained in the following. We can thereby confirm that the assumed Hopfield fits are consistent with our experimental findings and general graphene cyclotron absorption properties. This indicates the absence of LP softening, which would precede the SRPT in question.

The models put forward for graphene Landau polaritons so far \cite{chirolli_drude_2012,hagenmuller_cavity_2012} consider wavelength-sized Fabry-Pérot cavities, where the electromagnetic field is well described by a transverse vector potential $\vec{A}$. In the sub-wavelength setup considered here, graphene Landau excitations couple to the longitudinal near field of the complementary split ring resonator described in Coulomb gauge by the quasi-static scalar potential instead. To account for this difference, we derive a Hopfield Hamiltonian - which we show to be equivalent to that of Ref.~\cite{chirolli_drude_2012} - starting from the quasi-static scalar potential of the longitudinal cavity field coupled to the charge density of the graphene Landau excitations in the multipolar coupling scheme and neglecting fully static cavity-mediated electron-electron interactions. We find a $P^2$ term (that translates to a $A^2$ term via a Power-Zienau-Woolley transformation \cite{cohen_tannoudji_atom_photon_1998}) by tracing out all off-resonant field modes \cite{de2018cavity}, while coupling of the field to interband transitions is shown to be suppressed in the multipolar gauge (see Supplemental Material). This framework not only clarifies the origin of the $A^2$ term for the near-field cavity employed here, but also allows us to simulate the light-matter coupling constant via a finite-element simulation of the effective mode volume. The resulting computed normalized coupling ratios reported in Figure \ref{fig:fig3_fitting}b (solid line) show good agreement with our data, without any fitting parameters. In contrast, if we compare our data strengths to the prediction for Fabry-Pérot cavities \cite{chirolli_drude_2012,hagenmuller_cavity_2012}, we consistently find that these theories underestimate the observed coupling strength. This finding highlights the need for a theory that takes into account the sub-wavelength nature of the resonator as well as the ability of these resonators to enhance the light-matter coupling. 

In order to fully exclude the emergence of a Dicke SRPT in our system, we study the trends of the UP and LP dispersions as a function of density at specific detunings between the cavity and cyclotron frequency. We report the fitted peak positions of UP and LP normalized to the respective fitted cavity frequencies at the minimum magnetic field ($B$ = \qty{0}{T}), at maximum magnetic field ($B$ = \qty{6}{T}), and at resonance ($B = B_{\times}$) in Figure \ref{fig:fig4-Hopfield}a-c, respectively. All frequencies are plotted with error bars representing $\pm 3 \sigma$ derived from the fit. The shaded regions around the analytical curves indicate the spectral resolution of the measurements.

At $B$ = \qty{0}{T} (Figure \ref{fig:fig4-Hopfield}a), the UP frequency blue-shifts with increasing density, up to a factor of $1.2\times$ f$_{\text{cav}}$. This behavior is commensurate with polariton gap opening as a consequence of diamagnetic contributions \cite{chirolli_drude_2012, maissen_ultrastrong_2014}. These diamagnetic contributions $\Delta\omega$ (arising from the $\vec{A}^2$-term) distinguish the Hopfield and Dicke models, as they are neglected in a Dicke Hamiltonian \cite{hepp_superradiant_1973,wang_phase_1973,hagenmuller_cavity_2012,chirolli_drude_2012}.
The blue-shift can be expressed via the coupling strength \cite{maissen_ultrastrong_2014}
\begin{align}
    \frac{\Delta\omega}{\omega_\text{cav}} = \sqrt{4\eta^2 +1}
\end{align}
and shows good agreement with the data, as can be seen from the solid red line in Figure \ref{fig:fig4-Hopfield}a. In our sub-wavelength theory, the $P^2$ term has the same effect. In contrast, the UP mode of a Dicke model, which does not have any contributions from an $\vec{A}^2$-term, stays fixed to the cavity frequency, as shown by the gray dashed line in Figure \ref{fig:fig4-Hopfield}a.

An even starker contrast between the two theories is to be expected for the LP at $B=\SI{6}{T}$ (Figure \ref{fig:fig4-Hopfield}b). The hallmark of an SRPT is a continuous softening of the LP mode as the critical coupling is approached, as indicated by the gray line. We find a nearly constant LP frequency instead. While we observe a slight red-shift of $\approx$ \qty{5}{\%}, this can be explained by the change in the cyclotron slope ($B_\times$) with the density. This effects a gradual change in the asymptotic behavior. Effectively, at higher densities, the LP has not yet recovered to the bare cavity frequency at the maximum field of \SI{6}{T}. The softening of the LP mode in the Dicke model would induce a considerably stronger red-shift. For example, in the case of the strongest expected coupling strength, the LP mode would spectrally converge to a frequency as low as $\approx$ 0.4$\times f_{\text{cav}}$. In contrast, the calculated LP frequency in the Hopfield model (solid blue line, Figure \ref{fig:fig4-Hopfield}b) agrees with our experimental data. 

At the respective resonance point $B_\times$, we can study both UP and LP and directly observe the evolution of the branch splitting (Figure \ref{fig:fig4-Hopfield}c). Once more, the expectation for the Dicke branches is red-shifted considerably with regard to the measured data. The Hopfield branches adapted for the graphene system are overlaid as solid purple lines and again represent the data accurately. Selected corresponding spectra are shown in Figure \ref{fig:fig4-Hopfield}d. These data show the evolution of the polariton splitting as a function of density (coupling strength), represented by the offset of the y-axis of the normalized spectra. The corresponding $B_\times$ are denoted on the right hand side. The purple line represents the identical Hopfield dispersion plotted in Figure \ref{fig:fig4-Hopfield}c.

We therefore conclude that the behavior of both polariton branches, measured over a wide range of densities and detunings (controlled via the magnetic field), is consistently described by a Hopfield Hamiltonian and not the Dicke model. This proves the emergence of diamagnetic contributions, as argued in \cite{chirolli_drude_2012} and shown by our quasi-static model, and excludes the onset of a SRPT in graphene Landau polaritons. We note that the probed parameter space, fixed by the cavity mode $\omega_\text{cav}$, is predicted to undergo a phase transition at the critical coupling strength $\eta_{\text{c}} = 0.5$ and critical density $n_{\text{c}}$ = \qty{11.6E12}{cm^{-2}}, which is reachable in high-quality heterostructures \cite{zhou_isospin_2022}. While the measured range reported here lies below these critical values, the onset of this second order phase transition has been systematically disproved by investigating three distinct detuning regimes of the interaction. We also note that finite temperature effects at the experimental temperature of \qty{3}{K} do not appreciably alter the critical coupling condition, see Supplementary Figure S12.

The boundaries of the SRP are naturally affected by the chosen cavity frequency. For a cSRR resonant at f$_{\text{cav}}$ = \qty{500}{GHz}, the critical density decreases to $n_{\text{c}}$ = \qty{1.8E12}{cm^{-2}}. The normalized coupling strength, however, stays fixed to $\eta_{\text{c}} =0.5$ (see Supplementary Figure S13). A drawback in this scenario is the spatial increase of the cSRR size to accommodate the longer wavelength, necessitating an even larger exfoliated graphene flake. Fabrication, as well as graphene homogeneity, pose limitations in this case, though they are not of fundamental nature. For a phase diagram showcasing the dependence of the SRPT onset as a function of coupling strength and detuning, see Supplementary Figure S12.

\section{Conclusion}
We experimentally show that there is no onset of a SRPT in graphene Landau polaritons. We infer this from the absence of polariton softening as we increase the charge density in graphene. Our theoretical framework, adapted for sub-wavelength resonators, establishes the presence of a quadratic ($P^2$) term, akin to the $\vec{A}^2$-term in Fabry-Pérot cavity-coupling, that both ensures gauge invariance of the model and prevents a superradiant phase transition. Our findings thus highlight the importance of an accurate description of the cavity-field in the ultrastrong coupling regime, which seems particularly relevant in light of current studies of cavity-induced order of material degrees of freedom \cite{lu_cavity_2025}. The predicted occurrence of superradiant phase transitions in other Dirac materials and correlated materials seems unlikely \cite{liu_magnetopolariton_2014,li_ultrastrong_2016}. Yet, gauge-invariant theories predict a superradiant phase transition in the case of non-uniform cavity fields \cite{andolina-2020}. The required non-uniformity may be achieved in narrow-gap cSRRs, where we have shown that nonlocal effects can be probed \cite{rajabali_polaritonic_2021}. 

The presented results, together with those in \cite{helmrich_cavity-driven_nature_2026}, open the path for broadband spectroscopy experiments in the THz range with a multitude of different exfoliated van der Waals materials. This provides insights into the low-energy physics of such systems that have been shown to exhibit interaction-driven phenomena such as superconductivity \cite{cao_unconventional_2018}, Wigner crystallization \cite{Smolenski2021}, or fractional carrier statistics \cite{zengThermodynamicEvidenceFractional2023}. Such states exhibit collective modes at THz frequencies, which are routinely probed by spectroscopy in bulk quantum materials \cite{Basov2011}. For van der Waals materials, however, the inherent sub-wavelength nature of these materials poses a fundamental limit for far-field spectroscopy. This limitation is no longer present when using the solid immersion lens technique. In particular, sub-wavelength spectroscopy in a magnetic field allows for the extraction of the effective mass and thus provides insight into the band structure of a plethora of van der Waals materials. Excitingly, interaction-induced changes in the effective mass may be detected, allowing for quantitative parameter extraction in correlated states in a contact-free way.

\section{Acknowledgements}
The authors thank M. Barra Burillo for help in the initial part of the project. F.H. and T.F.N. thank A. \.{I}mamo\u{g}lu for his continuous support. E.J. thanks A. \.{I}mamo\u{g}lu for the use of his facilities.
We acknowledge the cleanroom facility FIRST at ETH Zürich. E.J. thanks G. Kipp for fruitful discussions. E.J. and F.H. acknowledge discussions on sample fabrication with the Ensslin group at ETH. F.L. acknowledges fruitful discussions with Johannes Feist. 

\section{Author Contributions}
E.J. fabricated the sample with help of F.H., J.F., and G.S. The terahertz generation source was provided by M.J. The terahertz set-up alignment and cryo-magnet maintenance were performed by E.J. and L.H. Optical measurements were performed by E.J. and L.H., with input from F.H., J.F., and G.S. Transport measurements were performed by E.J. and L.G., with input from F.H., J.F., and G.S. The data were interpreted by E.J., F.H., F.L., L.H., L.G., T.F.N., J.F., and G.S. The Hamiltonian model was developed by F.L. Finite element simulations were performed by E.J. with input from F.H., F.L., L.H., L.G., T.F.N., J.F., and G.S. The manuscript was written by E.J and F.H. with input from all authors. J.F and G.S. conceived and supervised the project.

\section{Competing interests}
The authors declare no competing interests.

\section{Funding}
E.J., L.H., L.G., J.F., and G.S. acknowledge funding by the Swiss National Science Foundation (SNF) (Grant numbers 10000397 and 200020-207795).
F.H. acknowledges support from the Swiss National Science Foundation (SNF) (Grant number  200020 207520). M.J. acknowledges funding by the US Department of Energy (Grant number DE-SC0016925). T.F.N. acknowledges funding from the ETH postdoctoral fellowship program. F.L. was supported by the Quantum Center Research Fellowship and the Dr Alfred and Flora Spälti Fonds. 

\bibliographystyle{naturemag}
\bibliography{bibliography}

\end{document}


\preprint{}

\title{Supplementary Information \linebreak Absence of a Superradiant Phase Transition in Dirac Landau Polaritons}

\date{\today}

\author{Elsa Jöchl}
\affiliation{Institute for Quantum Electronics, ETH Zürich, Zürich 8093, Switzerland}
\affiliation{Quantum Center, ETH Zürich, Zürich 8093, Switzerland}
\author{Felix Helmrich}
\affiliation{Institute for Quantum Electronics, ETH Zürich, Zürich 8093, Switzerland}
\affiliation{Quantum Center, ETH Zürich, Zürich 8093, Switzerland}
\author{Frieder Lindel}
\affiliation{Quantum Center, ETH Zürich, Zürich 8093, Switzerland}
\affiliation{Institute for Theoretical Physics, ETH Zürich, Zürich 8093, Switzerland}
\author{Lucy Hale}
\author{Lorenzo Graziotto}
\affiliation{Institute for Quantum Electronics, ETH Zürich, Zürich 8093, Switzerland}
\affiliation{Quantum Center, ETH Zürich, Zürich 8093, Switzerland}
\author{Mona Jarrahi}
\affiliation{Terahertz Electronics Laboratory, UCLA, Los Angeles 90095, United States}
\author{Tobia F. Nova}
\affiliation{Department of Quantum Matter Physics, University of Geneva, Geneva 1211, Switzerland}
\author{Jérôme Faist}
\affiliation{Institute for Quantum Electronics, ETH Zürich, Zürich 8093, Switzerland}
\affiliation{Quantum Center, ETH Zürich, Zürich 8093, Switzerland}
\author{Giacomo Scalari}
\affiliation{Institute for Quantum Electronics, ETH Zürich, Zürich 8093, Switzerland}
\affiliation{Quantum Center, ETH Zürich, Zürich 8093, Switzerland}

\maketitle
\tableofcontents
\newpage
\section{Sample Fabrication}
The fabrication protocol follows \cite{helmrich_cavity-driven_nature_2026}. Monolayer graphene and thin hBN flakes were exfoliated from bulk crystals using the scotch-tape method onto SiO$_2$/Si substrates and identified using an optical microscope. The thickness of the flakes was determined using the optical contrast with respect to the substrate. The thickness of hBN flakes was further confirmed by atomic force microscopy. We find \SI{33}{\nano\meter} for the top hBN and \SI{25}{\nano\meter} for the back hBN. The back gate was prepatterned using optical lithography and electron beam evaporation of \SI{4}{\nano\meter} Cr/Pt onto high-resistivity Si substrates with a 285 nm oxide layer in a lift-off process. Optical lithography was performed using direct laser writing of an image-reversal photoresist. The gate was contacted by gold leads patterned by another round of optical lithography and electron-beam evaporation. The van der Waals heterostructure was assembled from exfoliated flakes using a polymer dry-stacking technique employing a dome-shaped  polydimethylsiloxane (PDMS)/bisphenol-A polycarbonate (PC) stamp in an Ar-filled glove box. The heterostructure was released onto the back gate by melting the PC sacrificial layer. Residual PC was dissolved by immersion in chloroform. The monolayer graphene flakes were contacted using the edge-contact method \cite{wang_one-dimensional_2013}. The pattern was created using optical lithography; hBN was etched by reactive-ion etching using a combination of CHF$_3$ and O$_2$ gases, and \SI{10}{\nano\meter}/\SI{60}{\nano\meter} Cr/Au was deposited using electron beam evaporation. The same mask was used for the etching and deposition steps. Subsequently, a \SI{140}{\nano\meter} dielectric layer of alumina was grown on the full substrate using atomic layer deposition. The bond pads were subsequently opened again using a wet etching method with an Al etchant solution. Next, the resonator was fabricated using optical lithography and electron-beam evaporation of \SI{10}{\nano\meter}/\SI{190}{\nano\meter} Ti/Au. Finally, a \SI{2}{\micro \meter} protective layer of THz-transparent polymer (BCB, benzocyclobutene) was spin-coated onto the chip and cured at temperature of \SI{250}{\celsius} in a nitrogen atmosphere. The sample was mounted in a custom sample holder with the Si immersion lenses and wire-bonded using Au wires.
\newpage
\section{Terahertz Spectroscopy}

The data are obtained by way of THz time-domain spectroscopy (TDS) in a transmission setup. A NIR Titanium Sapphire laser centered at $\lambda =$ \qty{800}{nm}, with a repetition rate of $f_\text{rep}$ = \qty{80}{MHz} and a pulse length of $t =$ \qty{70}{fs}, is used to pump a THz photoconductive antenna (PCA), which is presented in \cite{turan_impact_2017}. We apply an AC voltage at $f =$ \qty{15}{kHz} with a $V_{pp} =$ \qty{10}{V} and a rectangular shape to the PCA, which generates broadband THz pulses spanning a range of $\Delta f \approx$ \qtyrange{0.1}{2.3}{THz}. Using parabolic mirrors in a 2f-2f setup, this THz light is guided through the cryostat containing the sample to be detected at the output using an electro-optic sampling detection scheme. The detection material is a \qty{3}{mm} thick [110] ZnTe crystal.

\begin{Sfigure}[h!]
    \centering
    \includegraphics[width=1\linewidth]{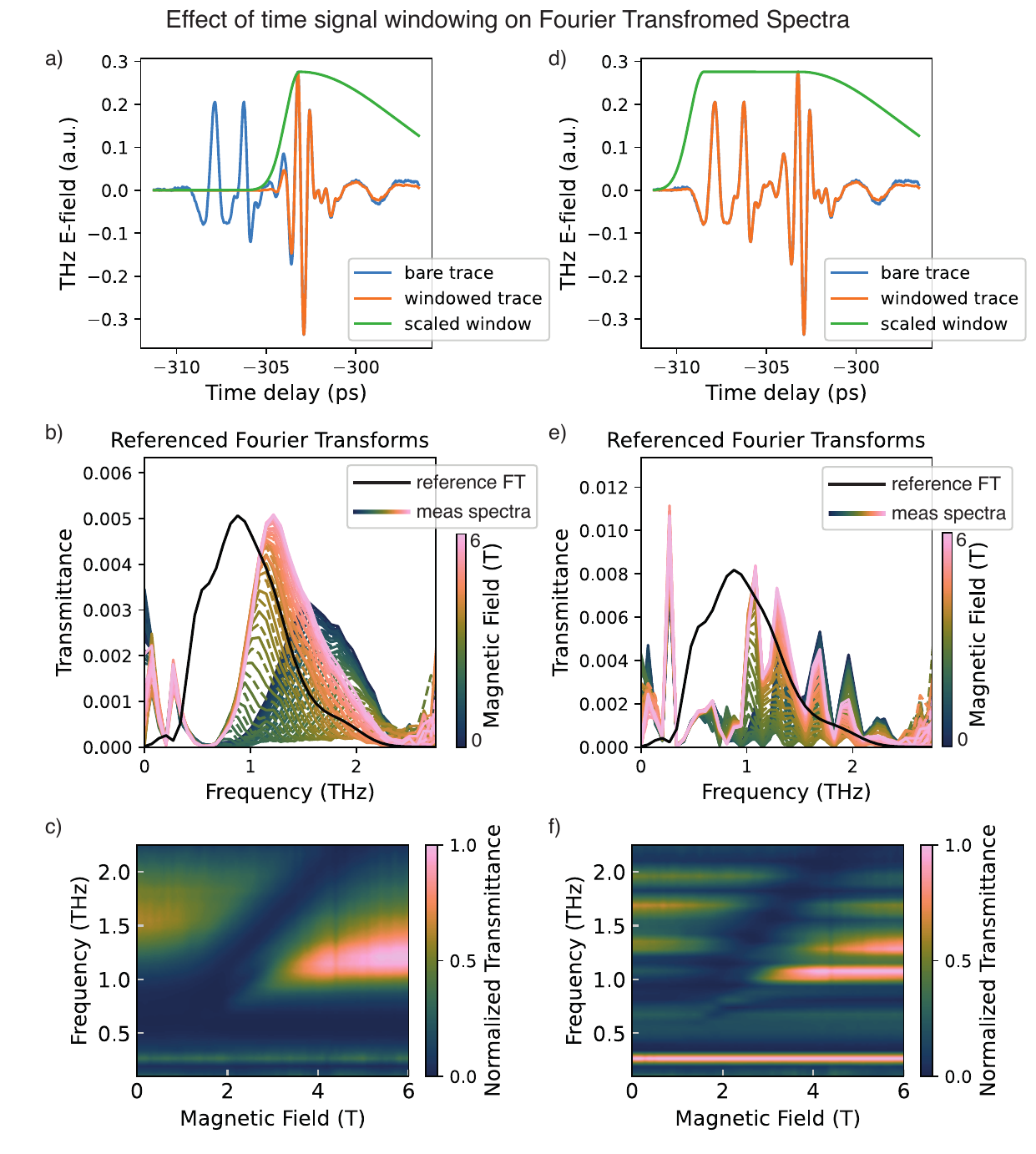}
    \caption{Experimental time domain data shown with the applied window function Fourier transforms and spectral maps according to the applied window function. Panels a-c) depict how all data presented and analyzed within this work were treated. d-f) Show the Fabry-Pérot fringes when including the first two echoes in the data analysis.}
    \label{supp-fig:windowing-timetraces}
\end{Sfigure}

The time traces obtained in this way include echoes that must be filtered out to avoid Fabry-Pérot fringes in the frequency domain. This was done by applying an asymmetric Gaussian window function to the bare time trace, as shown in Figure \ref{supp-fig:windowing-timetraces}a. Applying window functions in the time domain is common practice in THz-TDS analysis. Such echoes may result from the misalignment of the aSILs with respect to the sample chip. Panels \ref{supp-fig:windowing-timetraces}a-c show the procedure applied to all presented data. Smoothing the signal in the time domain reduces fringing effects arising from Fabry-Pérot reflections, enabling the analysis of smooth polariton dispersions. Panels \ref{supp-fig:windowing-timetraces}d-f present how including the pre-pulses introduces fringing while still showing transmission enhancement in the frequency range of the polariton branches. Furthermore, the pre-pulses exhibited no dependence on the magnetic field, supporting the argument that they arise from alignment or setup artifacts rather than physical effects related to the graphene sample.

All the resulting measured spectra are reported in Figure \ref{supp-fig:meas-spectra}, where the highest measured density is situated at the top right, and the lowest measured density at the bottom right of the Figure. The cyclotron slope is becoming steeper as the density is decreased, which shifts the anti-crossing magnetic field to lower magnetic field values. We further observe a change in the UP asymptotic frequency at $B=\SI{0}{T}$, which appears more blue-shifted for high densities (shallow cyclotron slopes).

\begin{Sfigure}[h]
    \centering
    \includegraphics[width=1\linewidth]{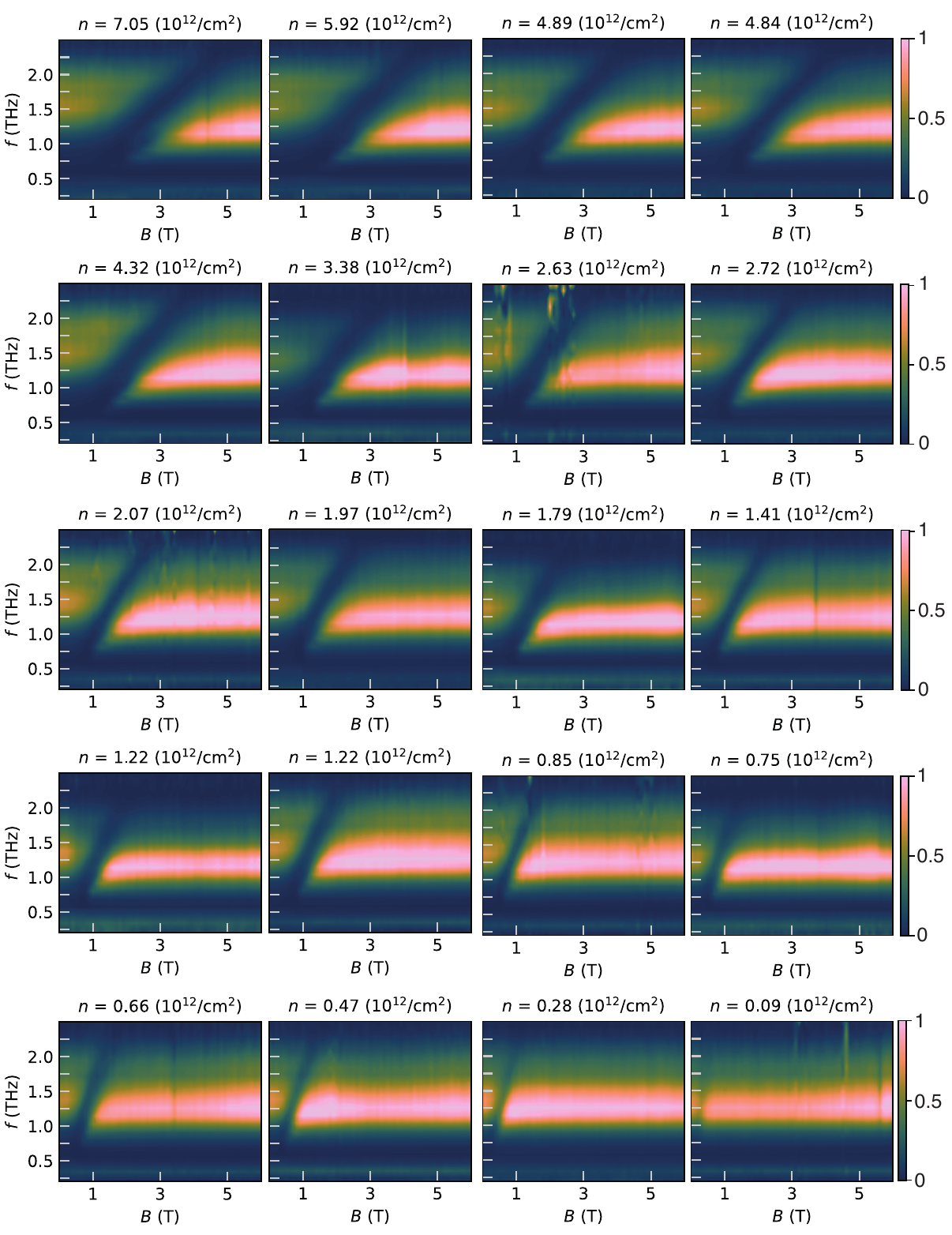}
    \caption{All measured spectra, normalized to their respective maxima to enhance contrast. The density (cyclotron slope) is decreasing (increasing) from left to right and top to bottom. }
    \label{supp-fig:meas-spectra}
\end{Sfigure}
\newpage

\section{Landau Level Spectrum in Monolayer Graphene}
\subsection{Theory}
As given in the main body of the text, the Landau Level (LL) energy of a LL with index $N$ in graphene is defined as the following:
\begin{align}\label{eq1}
    E_n = \text{sgn}(N) \sqrt{2 v_F^2 e  \hbar B(\nu)}\,\sqrt{N},
\end{align}
with the Fermi velocity $v_F \approx $ \qty{10E6}{m/s}, the electron charge $e$, the reduced Planck constant $\hbar$, and the magnetic field $B(\nu)$ as a function of the filling factor $\nu$. The selection rules allow for inter- and intra-band transitions according to $\vert N\vert \rightarrow \vert N \pm 1\vert$. We further define the filling factor in graphene as
\begin{align}\label{eq:nu}
    \nu = \frac{nh}{g_DeB(\nu)},
\end{align}
wherein $n$ is the carrier density, $h$ is the Planck constant, and $g_D = 4$ introduces the four-fold spin and valley degeneracy. Therefore, the considered LLs we account for are degenerate for both spin up and down states, as well as the $K$ and $K'$ valleys (this constitutes a negligible difference to the usual convention). As we expect not to be sensitive to spin- and valley-split states in spectroscopic measurements at $T =$ \qty{3}{K}, this description should fully encompass the transitions at hand. Cyclotron transitions between the LLs $N=\nu-\frac{1}{2} $ and $N+1 = \nu+\frac{1}{2}$ (relevant at THz frequencies) can now be considered and simplified for the case of high filling factors $\nu \gg 1$:
\begin{align}
    \hbar \omega_{\text{cyc}} &= \sqrt{2v_F^2\hbar eB} \,\Big(\sqrt{\nu +\frac{1}{2}}-\sqrt{\nu-\frac{1}{2}} \Big)  \\
     &= \sqrt{2v_F^2\hbar eB} \,\sqrt{\nu} \,\Big(\sqrt{1+\frac{1}{2\nu}}-\sqrt{1-\frac{1}{2\nu}} \Big) \\
     &\approx \sqrt{2v_F^2\hbar eB} \,\sqrt{\nu} \,\Big((1+\frac{1}{4\nu})-(1-\frac{1}{4\nu}) \Big) \\
     &= \sqrt{2v_F^2\hbar eB} \, \frac{1}{2\sqrt{\nu}} \label{eq:bbbb} \\
     &= \hbar \frac{eBv_F^2}{E_F} = \hbar \frac{eB}{m^*} \label{eq:wcyc} ,
\end{align}
where we defined the effective cyclotron mass of graphene carriers $m^* = E_F/v_F^2$. From \ref{eq:bbbb} we can infer the dependence of $B$ on $\nu$
\begin{align} \label{eq:Bnu}
    B(\nu) &= \frac{2\hbar \omega_{\text{cyc}}^2}{v_F^2e}\, \nu
\end{align}
Plugging Equation \ref{eq:Bnu} into Equation \ref{eq:nu} allows us to calculate how the filling factor depends on the carrier density
\begin{align}
    \nu &= \frac{n h e v_F^2 }{2\hbar g_D e\omega_{\text{cyc}}^2\nu } \\
    \nu&= \pm \sqrt{ \frac{\pi v_F^2}{ g_D\omega_{\text{cyc}}^2} \cdot n},
\end{align}
where the sign of the filling factor is given by electron (+) or hole (-) doping, respectively.

For resonant coupling between the cyclotron transition and the cavity mode, we impose $\omega_{\text{cyc}} = \omega_{\text{cav}}$. This enables us to calculate the anticrossing magnetic field:
\begin{align}
    B_{\times}(\nu) &= \frac{2\hbar \omega_{\text{cav}}^2}{v_F^2e}\, \nu
\end{align}

The filling factor at the anticrossing can then be explicitly expressed via the carrier density $n$:

\begin{equation}
    \nu_\times = \sqrt{ \frac{\pi v_F^2}{g_D \omega_{\text{cav}}^2}\cdot n}
    \label{eq:3}
\end{equation}




\subsection{AC Transport measurements}
In a three point contact transport measurement, we can verify the existence of Quantum Hall signatures in the system, as shown in Figure \ref{supp-fig:QH-meas}. The sample is placed in the same cryomagnet system as for the optical measurements, and cooled to the same temperature of $T=$ \SI{3}{K}. The contact scheme is overlaid in panel a of Figure \ref{supp-fig:QH-meas}. The red area in Figure \ref{supp-fig:QH-meas}a depicts the dimensions of the graphene flake, with the boundaries highlighted with a black dashed line. The unlabeled contact is physically disconnected from the flake to circumvent parasitic shorting to other areas on the sample and was therefore not utilized. We present transport data obtained by a lock-in detection scheme, with an applied voltage of \SI{100}{\milli \volt} across the graphene flake connected in series with a \SI{100}{\mega \ohm} resistor, resulting in a driven current of \SI{1}{\nano \ampere}. The signal is fed into a low-noise pre-amplifier, and detected at a modulation frequency of \SI{13.3}{Hz}. Then, a finite magnetic field is applied, and the source-drain voltage between contacts 1-2, as well as the voltage drop-off between contacts 1-3, is recorded as a function of the applied back-gate voltage $V_\text{BG}$ (carrier density). The measured voltages are converted into resistances and reported in Figure\ref{supp-fig:QH-meas} panels b and c as a function of the applied back-gate voltage ($V_\text{BG}$) with respect to charge neutrality at $V_{CNP}$. The charge neutrality voltage subtracted in these AC measurements is equal to $V_\text{CNP}=$ \SI{0.6}{V}. This was estimated via the asymmetry of the voltage measured across contacts 1-3, as will be discussed in the following.

The measured source-drain voltage $V_{12} = V_\text{SD}$ will be symmetric for both electrons and holes, and contain both static contact resistances $R_1$ and $R_2$ and a superposition of the longitudinal and transverse resistances, $R_{xx}$ and $R_{xy}$ respectively:
\begin{align}
    V_\text{SD} = I R_\text{SD} = I(R_1 + R_2 + R_{xy} + R_{xx})
\end{align}

We infer the contact resistances $R_1=$ \SI{1}{\kilo \ohm} and $R_2=$ \SI{0.67}{\kilo \ohm} and have subtracted them from the respective curves. For a fixed magnetic field direction and current direction, the voltage measured between contacts 1-3 will contain different contributions from electrons and holes:
\begin{align}
    V_{13}^{\text{electrons}} & = I R_{13} = I(R_1 + R_{xy}) \\
    V_{13}^{\text{holes}} & =  I R_{13} = I(R_1 + R_{xx})
\end{align}

This follows from considerations of the potential landscape following the asymmetric nature of the contact scheme. To gain insights on the longitudinal quantized resistance $R_{xx}$ for electrons, we have opted to subtract the resistance $R_{13}$ from the source-drain resistance $R_\text{SD}$. We note that this is only an approximate technique to verify oscillations of the same periodicity as the resistance $R_{xx}$ measured directly for hole doping. These symmetric oscillations can be seen for both measurements at $B=$\SI{2}{T} (Figure \ref{supp-fig:QH-meas}b, bottom) and at $B=$\SI{6}{T} (Figure \ref{supp-fig:QH-meas}c, bottom). Additionally, we note that the carrier mobility increases with increasing magnetic field, which is a possible reason for the increase in resistance between the two measurements. We infer from these traces that the quality of the sample is sufficient to exhibit Landau level discretizations, which were also verified by our optical transmission spectra. We will not discuss the physics of quantum Hall transport of graphene affected by the cavity vacuum field of the resonator \cite{xue2025observationcavitymediatednonlinearlandau}, due to constraints coming from contact and flake geometry, as well as lack of a reference Hall bar.

\begin{Sfigure}
    \centering
    \includegraphics[width=0.9\linewidth]{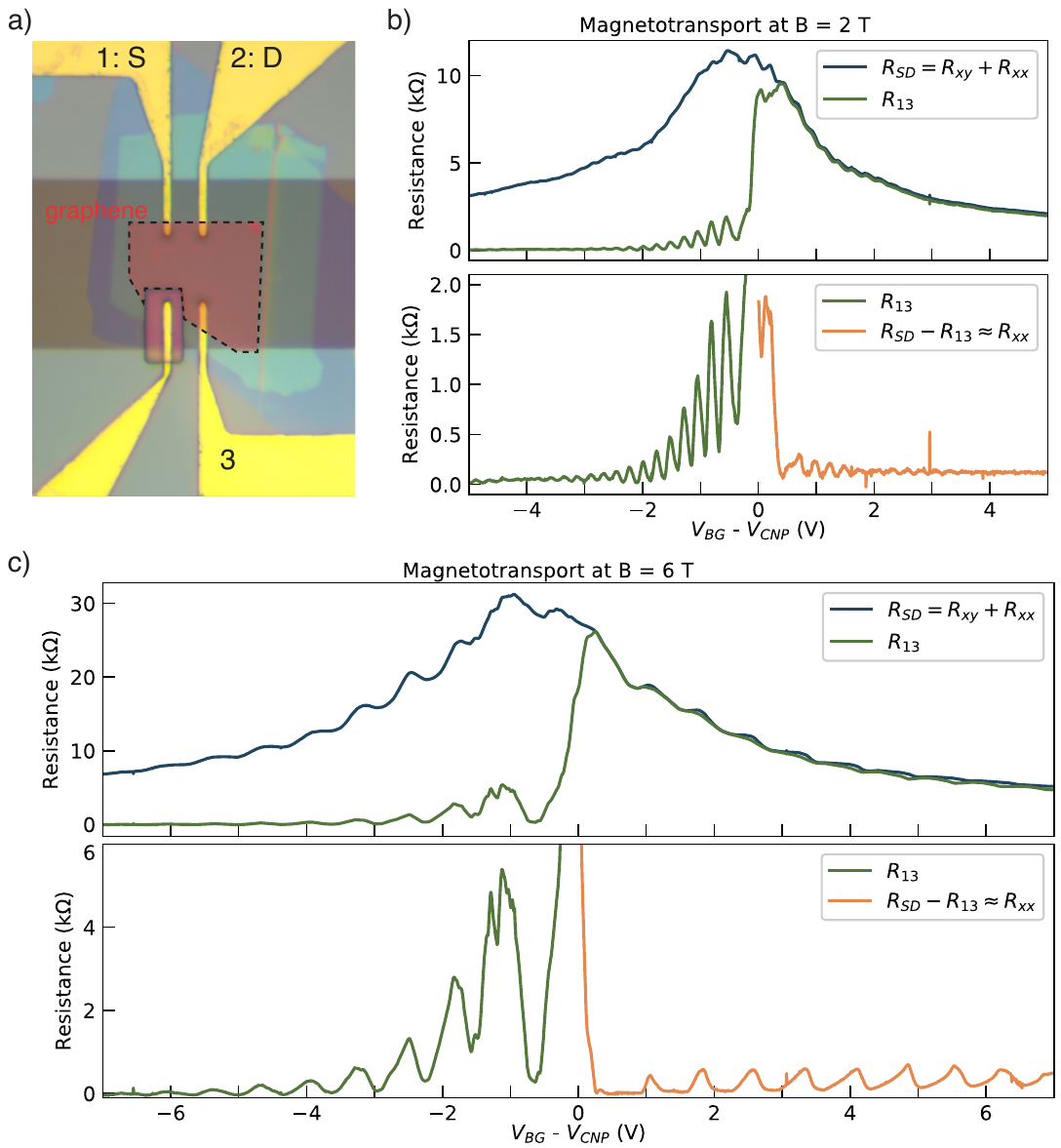}
    \caption{a) Contact scheme for the QH transport measurements. b) Measured magnetoresistance at B = \qty{2}{T} and c) at B = \qty{6}{T}.}
    \label{supp-fig:QH-meas}
\end{Sfigure}

\subsection{DC Transport measurements}
The DC transport behavior of the flake has also been characterized. For this, the sample is once more placed in the same cryomagnet system, and a Keithley source-meter is used to apply a finite DC bias ($V_{\text{gra}}$ = \SI{100}{\micro \volt}) to the flake with respect to ground. In this setting, one of the contacts shown in \ref{supp-fig:QH-meas}a is placed at this finite potential (source), while the other two are set to ground potential (drain). We then apply a voltage bias to the back-gate and sweep it across a wide range. In this way, we can measure the source-drain current flowing across the flake as a function of the applied voltage, see Figure \ref{supp-fig:transport1}a. We observe a stark change in the current just below $V_\text{BG}$ \SI{0}{\volt}. This curve does not represent a clean charge neutrality point behavior, as the current switches abruptly from negative to high positive values, continues to oscillate, and eventually settles at a slightly higher value on the electron side than on the hole side. This behavior indicates that there are finite potential barriers at the edge contacts, which lead to diode effects in the DC transport. Therefore, we repeat this measurement with the source placed on another contact. We take these measurements with varying magnetic fields in order to observe oscillations stemming from the LL quantization. 
The obtained data are shown in Figure \ref{supp-fig:transport1}b and c, with the bare traces shown separately in panel b, and the Landau fan depicted as a map in panel c. We again observe oscillations close to CNP just below $V_\text{BG}$ \SI{0}{\volt}. In the following supplementary section, we will present a way to estimate the $V_\text{CNP}$ from our spectroscopic measurements, which will show good agreement with the transport data that remains slightly obscured by diode effects.

\begin{Sfigure}[h!]
    \centering
    \includegraphics[width = \linewidth]{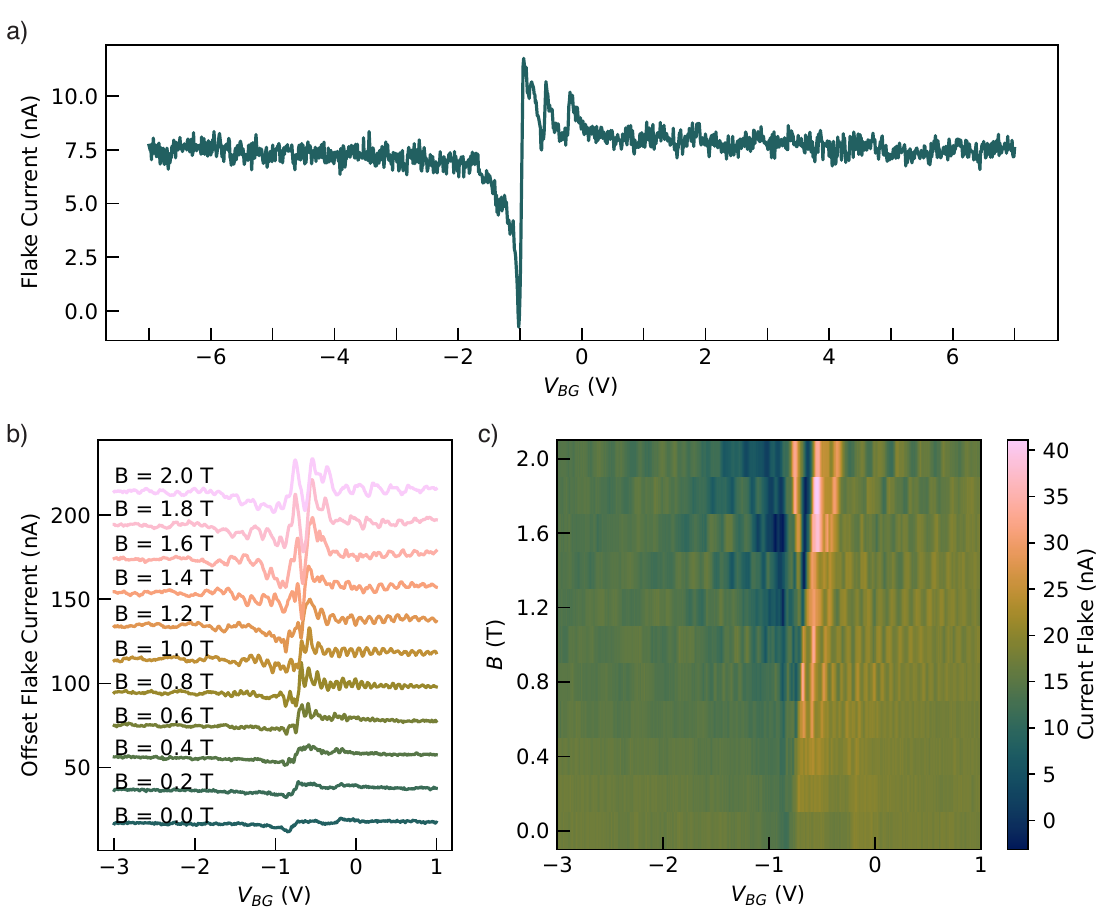}
    \caption{a) DC source-drain current measurement across the graphene flake as a function of applied back-gate voltage $V_\text{BG}$ containing diode effects around the CNP just below $V_\text{BG}$ \SI{0}{\volt}. b) The same measurement performed with the source placed on a different contact and for varying magnetic field strengths. c) The Landau Fan measured in panel b, depicted here as a current map.}
    \label{supp-fig:transport1}
\end{Sfigure}
\newpage
\section{Hopfield Model Fitting of measured spectra}
\begin{Sfigure}
    \centering
    \includegraphics[width=1\linewidth]{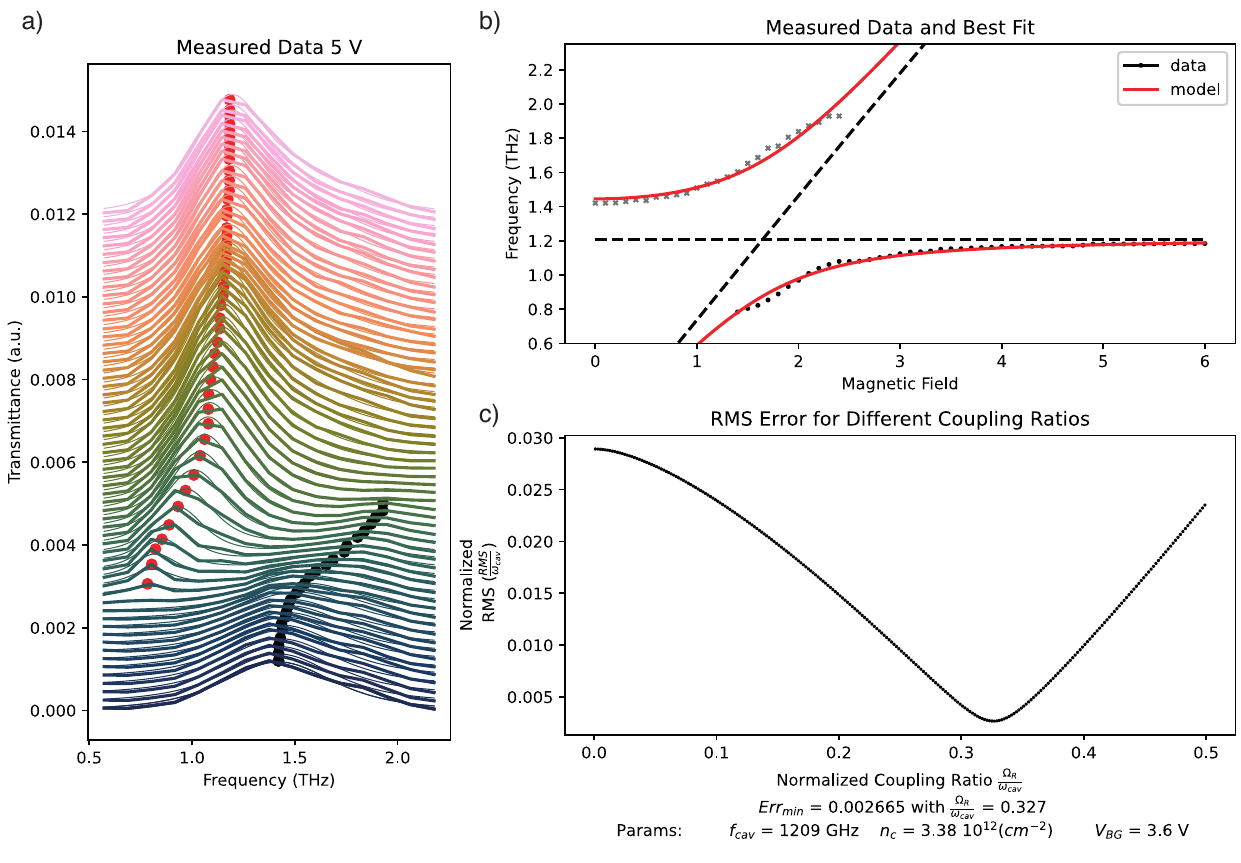}
    \caption{a) Lorentz curve fitting of the raw spectra, with overlaid extracted peak positions. b-c) Subsequent normalized RMS error estimation for different coupling strengths for a measurement at carrier density n $\approx$ \qty{3.6E12}{cm^{-2}}. These fits were systematically repeated with different parameter combinations of cavity frequencies and densities for each spectrum to obtain the global minimal error of each spectrum.}
    \label{supp-fig:fitting-procedure}
\end{Sfigure}
To determine whether the measured spectra are in agreement with the Hopfield model, we perform systematic fits as described in the main manuscript. In Figure \ref{supp-fig:fitting-procedure} we show such a fitting procedure for one selected measurement at an applied back-gate voltage of $V_\text{BG}=$ \SI{5}{V}. First, as shown in panel a, we fit the raw data with Lorentz functions - depending on magnetic field and presence of UP and/or LP we select a superposition of one or two Lorentz functions centered around the polariton center frequencies. The obtained peak frequencies are then collected and reported as a function of magnetic field, shown in panel b. The analytical Hopfield model is then fitted to these extracted points, and explicitly calculated for a range of cavity frequencies (slight deviations may arise from the spectral resolution, temperature, alignment, and their comparison to the shared reference measurement), and a range of densities (to dynamically fit the cyclotron frequency slope). The optimal coupling strength representing the extracted data points is then estimated by explicit calculation of the root mean square (RMS) deviation of each point from the model, as depicted in panel c. The overall minimal RMS error delivers the best fitting parameters.

We report all measured spectra, the extracted peak positions of both Lorentz-fitted LP and UP and their optimal fits with the Hopfield model polariton dispersion are shown in Figure \ref{supp-fig:fitting-spectra}. The measurements are ordered in the same fashion as Figure \ref{supp-fig:meas-spectra} above. We see that all measurements are consistently well represented by the Hopfield model.

\begin{Sfigure}[ht!]
    \centering
    \includegraphics[width=0.98\linewidth]{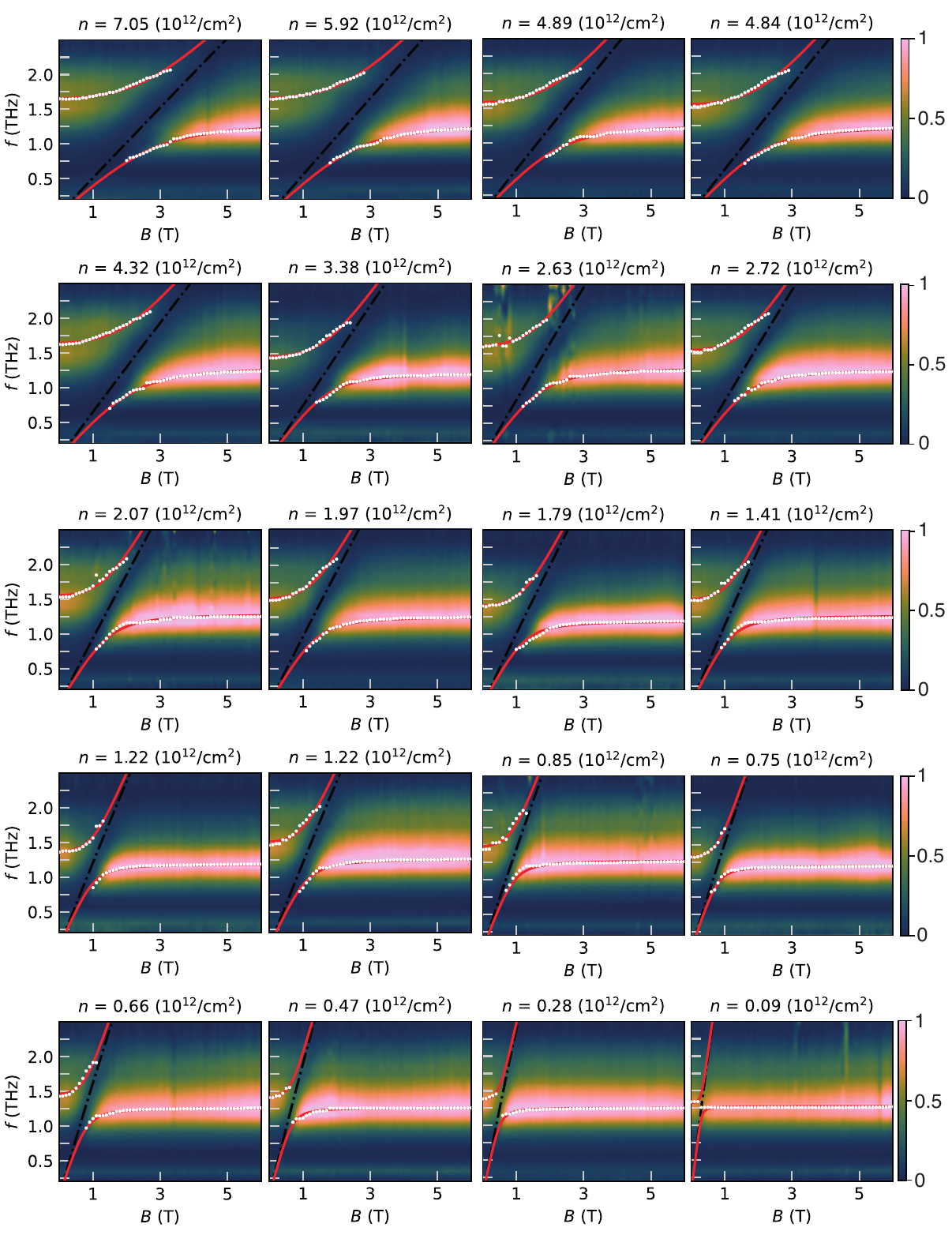}
    \caption{All measured spectra overlaid with the respective extracted peak positions and optimal fitted polariton dispersions according to the Hopfield model. The density (cyclotron slope) is decreasing (increasing) from left to right and top to bottom.}
    \label{supp-fig:fitting-spectra}
\end{Sfigure}

In Figure \ref{supp-fig:vappl-n}, we report the charge densities inferred from the Hopfield fits of the spectral maps and plot them as a function of applied back-gate voltage. We fit the slope of this linear increase in density with a conventional plate-capacitor model $$n= \frac{\varepsilon_0 \varepsilon_{\text{hBN}}}{e d_{\text{hBN}}}(V-V_{\text{CNP}}) .$$

A fit with this function yields a thickness of hBN $d_{\text{hBN}} \approx$ \qty{29}{nm}. This fitted hBN thickness is roughly commensurate with no-contact AFM measurements (yielding values around d = \qty{25}{nm}). The charge neutrality point, fitted at $V_{\text{CNP}}$ = \qty{-0.46}{V} is commensurate with the AC and DC transport measurements detailed above, which were affected by finite contact and diode resistances. The slight mismatch between holes and electrons is likely arising due to those same diode effects in the contacts.

\begin{Sfigure}
    \centering
    \includegraphics[width=0.75\linewidth]{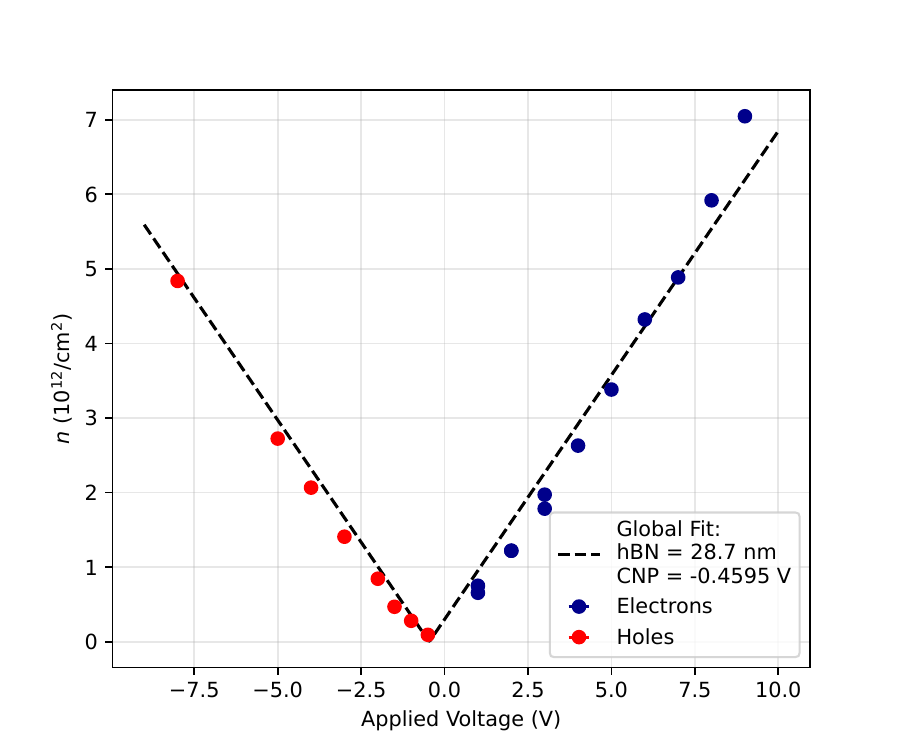}
    \caption{Fitted charge capacitor slope for carrier densities inferred from cyclotron absorption slopes as a function of the applied back-gate voltage.}
    \label{supp-fig:vappl-n}
\end{Sfigure}

\subsection{Effective Cyclotron Mass of Graphene Carriers}
From the measured spectra, we can infer the effective mass of carriers in graphene under the influence of strong magnetic fields. In Equation \ref{eq:wcyc}, we introduce an approximation for estimating the effective cyclotron mass of graphene by simplifying the cyclotron absorption energy $\hbar \omega_{\text{cyc}}$ for large filling factors $\nu \gg 1$. This approximation is overlaid with the explicit calculation of the respective LL transitions as a function of magnetic fields in Figure \ref{supp-fig:meff}a and shows perfect agreement with the explicit calculation in the measured frequency range $f<$ \SI{3}{\tera \hertz}. We can therefore analytically compute the effective mass in graphene as a function of the density via the approximation $m^* = E_{\text{F}}/v_{\text{F}}^2$ to good agreement with our data, and depict the analytical square-root dependence of $m^*$ on the carrier density $n$, as shown in Figure \ref{supp-fig:meff}b.

\begin{Sfigure}[ht!]
    \centering
    \includegraphics[width=0.95\linewidth]{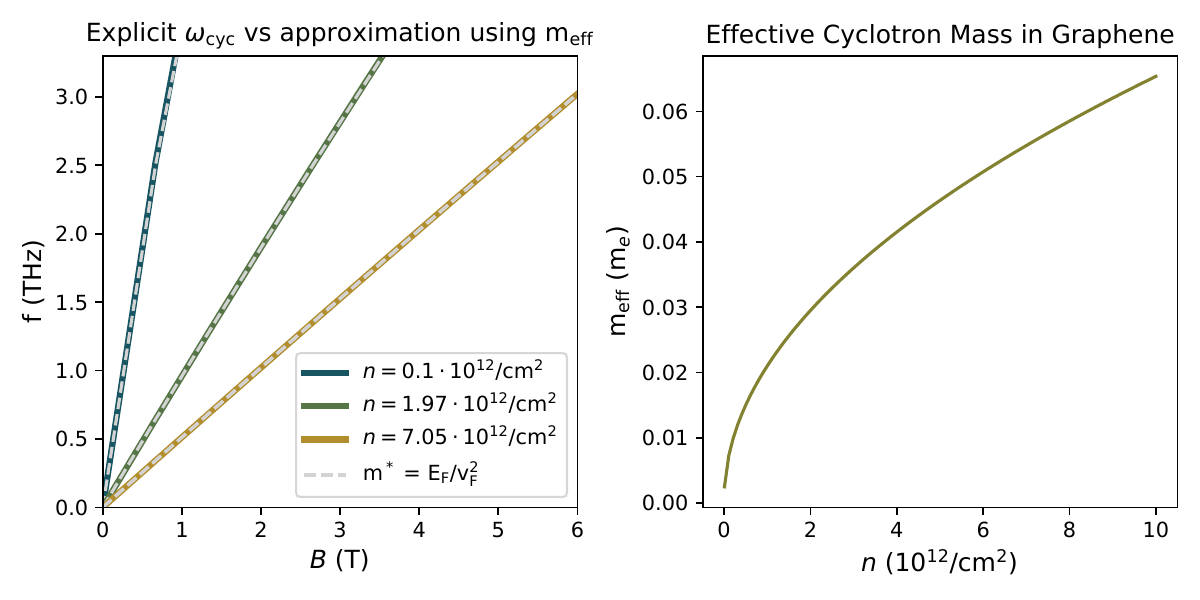}
    \caption{a) Calculation of the cyclotron absorption slope for the three different measurements presented in detail in the main manuscript at low, intermediate and high densites. Each curve is overlaid with their respective approximation of the cyclotron absorption slope calculated via the Fermi energy $m^* = E_{\text{F}}/v_{\text{F}}^2$. b) Analytic calculation of the effective cyclotron mass of graphene (in units of the electron rest mass $m_\text{e}$) in accordance with the measured spectra.}
    \label{supp-fig:meff}
\end{Sfigure}
\newpage

\section{Lower Polariton Splitting}
Spectra measured at high carrier densities for both electron and hole doping (see spectra with $n>$ \qty{4}{10^{12}/cm^2} in Figure \ref{supp-fig:meas-spectra}) show a peculiar feature close to the anticrossing magnetic field $B_{\times}$. The LP exhibits a blue-shifted side peak, much lower in transmission amplitude than the main LP branch. In Figure \ref{supp-fig:lp-split} we show explicit spectra for these spectra around their respective $B_{\times}$. We depict the zero-padded Fourier-transformed data (blue lines). Zero-padding a given time-domain signal before Fourier-transforming can artificially increase the spectral resolution, and can pronounce features otherwise difficult to assess. To verify the positions of these observed side peaks, the bare data is overlapped with orange lines. This feature is most pronounced at the highest measured density, and evolves with increasing magnetic field. The origin of such a split LP is not trivially explained. 

\begin{Sfigure}[ht!]
    \centering
    \includegraphics[width=1\linewidth]{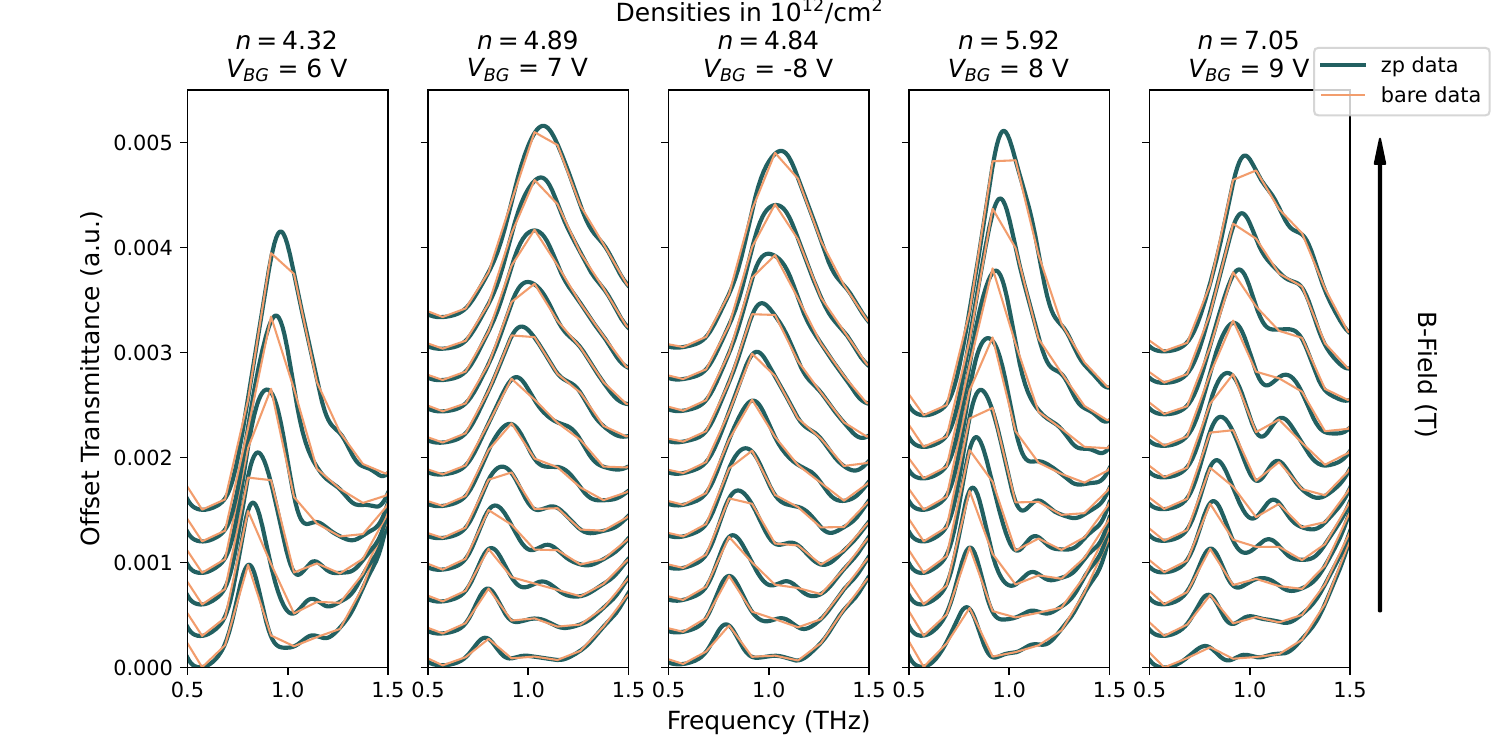}
    \caption{Transmittance spectra showing the splitting of the LP mode close to the respective anticrossing magnetic fields $B_{\times}$. The curves are y-offset with magnetic field values increasing in \SI{0.100}{\tesla} increments. The density increases from left to right, as depicted above. The green lines show zero-padded (zp) data, whereas the orange lines show the bare FT data.}
    \label{supp-fig:lp-split}
\end{Sfigure}

The origin of the splitting might arise from two separate plasmon modes interacting with the cavity mode centered at the same frequency. These distinct plasmon modes can be induced by the asymmetric shape of the graphene flake, as studied in detail in Ref. \cite{kipp_cavity_2025}. The lack of anti-crossing between the two LP modes suggests that their spectra are non-coherently overlapping. An interaction of this form could be explained by the interaction of the respective plasmon modes with the cavity at different momenta of the photonic field. Since the cavity field is dominated by electromagnetic contributions around $k\approx 0$, the low-momentum plasmon couples more strongly than that at finite momentum, leading to a splitting of the two LP branches. 

We would like to note, that the inclusion of a high-momentum photonic mode is moreover theorized to weaken the selection rule for dipolar LL transitions from $N \rightarrow N\pm 1$ to allow for quadrupole (and higher order) transitions $N \rightarrow N\pm 2$ \cite{bacciconi_theory_2025}. In this case, the same argument for the weaker coupling strength to the electromagnetic field at finite momentum holds as above.


\newpage
\section{Quasistatic model}
While previous theoretical models of Landau polaritons have focused on wavelength-sized cavities \cite{hagenmuller_cavity_2012,hagenmuller_ultrastrong_2010}, we here derive a model that takes the quasistatic (longitudinal) nature of the near-field resonator of the experiment into account. Note that this model should also be applicable to Landau polaritons in a semiconductor quantum well two-dimensional electron gas strongly coupled to a subwavelength resonator as in e.g.~Refs.~\cite{scalari_ultrastrong_2013,rajabali_polaritonic_2021,mornhinweg_mode-multiplexing_2024}.

We first derive a general single-mode light-matter Hamiltonian in the multipolar coupling scheme, before finding an appropriate expression for the charge density of the two-dimensional electron gas. This results in a multipolar Hopfield Hamiltonian for which we derive the coupling strength and discuss the role of interband transitions. 

\subsection{Single-mode reduction of the light matter coupling in the multipolar coupling scheme}

We start with the light-matter coupling Hamiltonian in the multipolar coupling scheme
\begin{align}
    H_\mathrm{dg} = H_F + H_\mathrm{LL}  + \frac{1}{2\epsilon_0} \int \dif^3 r  \vec{P}(\vec{r}) \cdot \vec{E}(\vec{r})  + \frac{1}{2\epsilon_0} \int \dif^2 r P^2(\vec{r}).
\end{align}
Here, $H_\mathrm{LL}$ and $H_\mathrm{F}$ are the bare electronic Landau level Hamiltonian including the interaction with the classical applied magnetic field and the bare Hamiltonian of the quantized field, respectively. Furthermore, $\vec{P}$ is the polarization field. The electric field operator in the multipolar coupling scheme $\vec{E}$ contains both longitudinal and transverse components. Importantly, the longitudinal components describe the quasistatic near field fluctuations of the subwavelength resonator, i.e. the dynamical longitudinal field generated by the charges inside the resonator. As we eventually only consider longitudinal electric-field modes that are unaffected by the Power–Zienau–Woolley (PZW) transformation, we note that a corresponding calculation in the minimal coupling scheme should yield identical results. 

Next, we single-out the lowest mode of the split-ring resonator with frequency $\omega_\mathrm{cav}$ given by 
\begin{align}
    \vec{E}_\mathrm{cav} = \sqrt{\frac{\hbar \omega_\mathrm{cav} }{2\epsilon_0 V_\mathrm{eff}}}  \vec{f}_\mathrm{cav}(\vec{r})a_\mathrm{cav} +  \mathrm{h.c.}.
\end{align}
Here, $\vec{f}_\mathrm{cav}(\vec{r})$ is the mode function, $a_\mathrm{cav}$ is a bosonic annihilation operator, and the effective mode volume is given by \footnote{Here, we neglected the evanescent field inside the metal. This, in general, would modify the mode volume. As the penetration depth in the THz into the metal is so low, we neglect this contribution here.}
\begin{align}
    V_\mathrm{eff} =  \int \dif^3r \, \epsilon(\vec{r}) |f_\mathrm{cav}(\vec{r})|^2,
\end{align}
where $\epsilon(\vec{r})$ is the permittivity of the cavity. 

Note that a full ab initio treatment, which could also identify and account for possible multimode resonances of the near-field resonator, could be derived based on the effective spectral density seen by the Landau system following Refs.~\cite{medina2021few,hale2025multi}. However, this goes beyond the scope of this work.

Next, we adiabatically eliminate all field modes except the singled-out mode $E_\mathrm{c}$ following Refs.~\cite{de2018cavity,sanchez2024general} by assuming $\omega_k\gg\omega_\mathrm{cyc}$ for all $k$ except for $k=\mathrm{cav}$. We first expand the electric field in modes
\begin{align}
    \vec{E}(\vec{r}) = \sum_k  \sqrt{\hbar \omega_k/(2\epsilon_0)} \overline{\vec{f}}_k(\vec{r}) a_k + \mathrm{h.c.}
\end{align}
where we used the normalization $\int \, \dif^3 r \overline{\vec{f}}_k(\vec{r}) \overline{\vec{f}}^\ast_{k^\prime}(\vec{r}) = \delta_{kk^\prime}$.
As a result, compare Appendix C of Ref.~\cite{de2018cavity}, we find
\begin{multline} \label{eq:Hdg_traced}
    H_\mathrm{dg} = \omega_\mathrm{cav} a_\mathrm{cav}^\dagger a_\mathrm{cav} + H_\mathrm{LL} + \sqrt{\frac{\hbar \omega_\mathrm{cav}}{2\epsilon_0 V_\mathrm{eff}}}(a_\mathrm{cav}+a_\mathrm{cav}^\dagger) \int \dif^3 r  \vec{P}(\vec{r}) \cdot \vec{f}_\mathrm{cav}(\vec{r})  + \frac{1}{2\epsilon_0} \left[\int \dif^3 r \vec{f}_\mathrm{cav}(\vec{r}) \cdot \vec{P}(\vec{r}) \right]^2 \\+  \frac{1}{2\epsilon_0} \int \dif^3 r \int \dif^3 r^\prime \vec{P}(\vec{r}) \cdot \boldsymbol{\lambda}(\vec{r},\vec{r}^\prime) \cdot \vec{P}(\vec{r}) ,
\end{multline}
with
\begin{align}\label{eq:mode_sum}
    \boldsymbol{\lambda}(\vec{r},\vec{r}^\prime) = \boldsymbol{\delta}(\vec{r}- \vec{r}^\prime) - \sum_k \overline{\vec{f}}_k(\vec{r}) \overline{\vec{f}}^\ast_k(\vec{r}^\prime).
\end{align} 
The last term in the first row of Eq.~\eqref{eq:Hdg_traced} avoids double counting the cavity mode that is not eliminated but still contained in the mode summation in Eq.~\eqref{eq:mode_sum}

As detailed in Ref.~\cite{sanchez2024general}, the mode summation in Eq.~\eqref{eq:mode_sum} can be done for general optical environments via macroscopic quantum electrodynamics by rewriting it in terms of the dyadic Green tensor $\mathbf{G}$ of the vector Helmholtz equation and using its properties in the complex plane. $\mathbf{G}$ accounts for the presence of the cavity via its complex permittivity. This procedure show that $ \boldsymbol{\lambda}(\vec{r},\vec{r}^\prime)$ yields the cavity-assisted Coulomb Kernel:
\begin{align}
    \boldsymbol{\lambda}(\vec{r},\vec{r}^\prime)  = \left[\frac{\omega^2}{c^2}\mathbf{G}(\vec{r}, \vec{r}^\prime, \omega)\right]_{\omega=0}.
\end{align}
The last term in Eq.~\eqref{eq:Hdg_traced} therefore describes the (fully not quasi-) static cavity-assisted electron-electron interaction. It contains free space-contributions, corresponding to the standard Coulomb interactions in graphene, as well as a cavity-induced contribution, describing e.g. static image charge effects. We will neglect the influence of both contributions in the following. If significant, such static electron-electron interactions can lead to ferroelectric instabilities, resulting in a softening of the lower polariton mode \cite{de2018cavity}. The absence of such a softening of the lower polariton in the experimental data implies the absence of such ferroelectric phase transitions in the current setup. A detailed analysis of the influence of static cavity-mediated electron-electron interactions in Dirac Landau polaritons will be subject of subsequent work. 

Next, we use that the cavity mode is longitudinal, which means that its mode function can be expressed in terms of a scalar potential $\varphi_c$: $\vec{f}_c = \nabla \varphi_c$ [$\varphi_c(\vec{q}) = -\mi \vec{q} \cdot \vec{f}_c(\vec{q})/q^2 $], such that 
\begin{align}
\int \dif^3 r  \vec{P}(\vec{r}) \cdot \vec{f}_c(\vec{r}) &  = -e\int \dif^3 r n(\vec{r}) \varphi_c(\vec{r}) ,
\end{align}
where we integrated by parts and used $\boldsymbol{\nabla} \cdot \vec{P} = -\rho$, where $\rho =- e n$ is the charge density of the graphene sheet and $n$ its electron density. So we see that the longitudinal field only couples to the charge density, i.e. the longitudinal polarization density :
 \begin{align} \label{eq:Pton}
    \vec{P}^{\parallel}(\vec{q}) = -e\frac{ \vec{q}}{q^2} n(\vec{q}).
\end{align}

\subsection{Charge density in graphene}

In this section, we derive the charge density operator in single-layer graphene. For a more detailed discussion, see e.g.~Ref.~\cite{goerbig2011electronic}. Because the charge density operator is not affected by the PWZ transformation, we can use results obtained via the minimal coupling scheme. 

We start with the bare Hamiltonian of the Landau levels 
\begin{align}
    H_\mathrm{LL} = \sum_{k_x,  n}  \epsilon_{n} c^\dagger_{k_x,n}c_{k_x n}
\end{align}
where $n = 0,\pm 1, \pm 2, \dots$, $\epsilon_n = \mathrm{sgn}(n) (\hbar v_F/l_B)\sqrt{2n}$ and $l_B = \sqrt{\hbar c/eB}$. Each Landau level has degeneracy $N_L  = g_D S /(2\pi l_B^2)$, where $g_D = 4$ accounts for spin and valley degeneracy and $S$ is the area of the 2DEG. 

Next, we define bright operators 
\begin{align} \label{eq:definition_bright}
    b_{n n^\prime}(\vec{q}_\parallel) = \frac{1}{\sqrt{N_L}} \sum_{k_x} \me^{\mi k_x q_y l_B^2} c^\dagger_{k_x-q_x,  n^\prime}c_{k_x, n}.
\end{align}
Note that they differ from previous definitions, in which the 2DEG was assumed quasi one dimensional fixing $q_y$ to the lowest fundamental mode \cite{hagenmuller_ultrastrong_2010}. 

Next, we approximate the 2DEG operators by their expectation value relative to the Fermi sea
\begin{align}
    \ket{\mathrm{GS}} = \Pi_{n  \in  N_\mathrm{occ}} \Pi_{k_x} c_{n  k_x}^\dagger \ket{0}.
\end{align}
Within this approximation, we have
\begin{align}
     c^\dagger_{k_x-q_x, n^\prime}c_{k_x^\prime-q_x^\prime, n^{\prime\prime\prime}} \approx \braket{\mathrm{FS}|   c^\dagger_{k_x-q_x, n^\prime}c_{k_x^\prime-q_x^\prime, , n^{\prime\prime\prime}} |\mathrm{FS}} = 0,
\end{align}
for $n^\prime, n^{\prime\prime \prime} \in N_\mathrm{unocc}$, and 
\begin{align}
     c^\dagger_{k_x,n}c_{k_x^\prime-q_x^\prime + q_x, n^{\prime\prime}} \approx \braket{\mathrm{FS}|   c^\dagger_{k_x, n}c_{k_x^\prime-q_x^\prime + q_x, , n^{\prime\prime}} |\mathrm{FS}} = \delta_{q_x^\prime q_x}  \delta_{nn^{\prime \prime}},
\end{align}
for $n, n^{\prime\prime} \in N_\mathrm{occ}$.
As a result, we find
\begin{align}
    [ b_{n n^\prime }(\vec{q}_\parallel), b_{n^{\prime\prime} n^{\prime\prime\prime} }^\dagger(\vec{q}_\parallel^\prime)] = \delta_{nn^{\prime\prime}}  \delta_{n^\prime n^{\prime\prime\prime}}  \delta_{q_xq_x^\prime} \delta_{q_y q_y^\prime},  
\end{align}
and the Landau level Hamiltonian can be expressed in terms of these bosonic creation and annihilation operators describing transitions $n^\prime \to n$ with frequency $\omega_{n,n^\prime} =\epsilon_{n}-\epsilon_{n^\prime }$:
\begin{align}
    H_\mathrm{LL} = \sum_{\substack{n' \in N_{\mathrm{unocc}} \\ 
    n \in N_{\mathrm{occ}}}}
\sum_{q_x, q_y} \omega_{n,n^\prime}  b^\dagger_{n n^\prime }(\vec{q}_\parallel)b_{n  n^\prime }(\vec{q}_\parallel)
\end{align}
The charge density can be written as
\begin{align} \label{eq:charge_density_general}
    n_{\vec{q}_{\parallel}} \approx 
\sum_{\substack{n' \in N_{\mathrm{unocc}} \\ 
n \in N_{\mathrm{occ}}}}
\sum_{k_x, k_x'}
\Big[
\langle k_x', n' | \hat{n}_{\vec{q}_{\parallel}} | k_x, n\rangle
\, \hat{c}^\dagger_{k_x', n'}
\hat{c}_{k_x, n}
+
\langle k_x, n | \hat{n}_{\vec{q}_{\parallel}} | k_x', n' \rangle
\, \hat{c}^\dagger_{k_x, n}
\hat{c}_{k_x', n'}
\Big]
\end{align}
The matrix elements read
\begin{align} \label{eq:n_matrix_elements}
\braket{ k_x', n'| \hat{n}_{\vec{q}_{\parallel}} |k_x, n }
= \me^{\mi q_y (2k_x - q_x) l_B^2/2}
e^{- q_{\parallel}^2 \ell_B^2 / 4}
\,
\delta_{k_x',\, k_x - q_x}
\,
F_{n',\, n}(\vec{q}_\parallel),
\end{align}
with \cite{goerbig2011electronic}
\begin{align}
   F_{n',n}(\vec{q}_\parallel)
=
\left[ C_{n'}^- C_n^-
F^{\mathrm{2DEG}}_{|n'|-1,|n|-1}(\vec{q}_\parallel)
+ C_{n'}^+ C_n^+
\mathrm{sgn}(nn^\prime)
F^{\mathrm{2DEG}}_{|n'|,|n|}(\vec{q}_\parallel)
\right],
\end{align}
and
\begin{align}
    F^{\mathrm{2DEG}}_{n',n}(\vec{q}_\parallel)
=
\sqrt{\frac{\min(n,n')!}{\max(n,n')!}}
\left(
\frac{(q_y + i q_x)\ell_B}{\sqrt{2}}
\right)^{|n'-n|}
L_{\min(n,n')}^{|n'-n|}
\!\left(\frac{q^2 \ell_B^2}{2}\right),
\end{align}
where $L_{n}^{n^\prime}$ are associated Laguerre polynomials and $C_n^\pm = \sqrt{(1\pm \delta_{n0})/2)}$. 

We can use Eqs.~\eqref{eq:n_matrix_elements} and \eqref{eq:definition_bright} in Eq.~\eqref{eq:charge_density_general}, to find
\begin{align}
     n_{\vec{q}_\parallel} 
   & =  \me^{\mi q_y q_x l_B^2/2  - q_\parallel^2 l_B^2/4 } \sqrt{N_L} \sum_{n,n^\prime} [F_{n,n^\prime}(\vec{q}_\parallel) b_{n,n^\prime}(\vec{q}_\parallel)+\mathrm{h.c.}].
\end{align}
Next, we take the continuum limit for which $b_{nn^\prime}(\vec{q}_\parallel) \to \sqrt{(2\pi)^2/S}b_{nn^\prime}(\vec{q}_\parallel)$ such that 
\begin{align}
    n(\vec{r}_\parallel) & = \frac{1}{2\pi \sqrt{S}}\int \dif^2 q_\parallel  \me^{\mi \vec{q}_\parallel \cdot \vec{r}_\parallel} \me^{\mi q_y q_x l_B^2/2  - q_\parallel^2 l_B^2/4 } \sqrt{N_L} \sum_{n,n^\prime} [F_{n,n^\prime}(\vec{q}_\parallel) b_{n,n^\prime}(\vec{q}_\parallel)+\mathrm{h.c.}],
\end{align}
where $n(\vec{r})\approx \delta(z-z_0)n(\vec{r}_\parallel)$ and $n, n^\prime$ correspond to occupied and unoccupied Landau levels, respectively.

In a last step, we note that the momentum $q_\parallel$ resolved by the cavity is mostly restricted by the gap size $d $ of the resonator to $q_\parallel < 1/d$ and therefore satisfies $q_\parallel l_B \ll 1$. In this limit and further assuming $\nu\gg1$ we have
\begin{align}
F_{ \nu - 1,\nu} =\frac{(q_y + \mi q_x)l_B}{\sqrt{2}} (C_{|\nu|-1}^- C_{|\nu-1|-1}^- \sqrt{\nu-1} + C_{|\nu|}^+C_{|\nu-1|}^+  \mathrm{sgn}(\nu \nu^\prime) \sqrt{\nu}),
\end{align}
where $\nu$ is the first unoccupied Landau level. For all other higher order transitions with $n^\prime \in N_\mathrm{unocc}$ and $n \in N_\mathrm{occ}$, e.g., for transition from $\nu-1$ to $\nu+1$, $\nu+2$ etc, we find that $F_{n^\prime n}$ is at least of order $q_\parallel^2 l_B^2$. To lowest order in $q_\parallel l_B$, we therefore have
\begin{align}
  n(\vec{r}_\parallel) = \sqrt{\frac{l_B^2 N_L}{8\pi^2 S}}\int \dif^2 q_\parallel (q_y + \mi q_x) (\sqrt{\nu-1} + \mathrm{sgn}(\nu \nu^\prime) \sqrt{\nu}) b_{\nu-1,\nu}(\vec{q}_\parallel)+\mathrm{h.c.}].
\end{align}

Only restricting to the resonant intraband transition for which $n,n^\prime >0$ and approximating $\nu\gg1$, such that $\sqrt{\nu} + \sqrt{\nu+1} \approx 2 \sqrt{\nu} $, we find 
\begin{align} \label{eq:n_intraband}
  n(\vec{r}_\parallel) = \sqrt{\frac{l_B^2 N_L \nu}{4\pi^2 S}}\int \dif^2 q_\parallel \me^{\mi \vec{q}_\parallel \cdot \vec{r}_\parallel} (q_y + \mi q_x)  b_{N-1,N}(\vec{q}_\parallel)+\mathrm{h.c.}.
\end{align}

\subsection{Hopfield model and coupling strength}

Inserting Eq.~\eqref{eq:Pton} and \eqref{eq:n_intraband} into Eq.~\eqref{eq:Hdg_traced} and neglecting the static electron-electron interactions, we find
\begin{align}
      H_\mathrm{dg} = \omega_\mathrm{cav} a_\mathrm{cav}^\dagger a_\mathrm{cav}+ H_\mathrm{LL}  + \hbar g (a_\mathrm{cav}+a_\mathrm{cav}^\dagger) (B + B^\dagger)  + \frac{\hbar g^2}{ \omega_\mathrm{cav}} (B + B^\dagger)^2 
\end{align}
where
\begin{align}
    B + B^\dagger = \sqrt{\frac{ \omega_\mathrm{cav} e^2}{2 \hbar \epsilon_0 V_\mathrm{eff}g^2}} \int \dif^3 r  \vec{P}(\vec{r}) \cdot \vec{f}_\mathrm{cav}(\vec{r}),
\end{align}
are the bright emitter operators. The coupling strenght $g$ has to be determined from the normalization condition $[B, B^\dagger] = 1$:
\begin{align}
   g^2 &=\frac{2 \pi^2 e^2\omega_\mathrm{cav} l_B^2 N_L \nu  }{ \hbar \epsilon_0 S  V_\mathrm{eff}} \frac{1}{(2\pi)^4 }\int \dif^2 q_\parallel \Big| \frac{ \vec{q}_\parallel \cdot \vec{f}_\mathrm{cav}(\vec{q}_\parallel)}{q_\parallel}\Big|^2 \\
    & = \frac{2\pi^2 e^2 \omega_\mathrm{cav} l_B^2 N_L  \nu}{ \hbar \epsilon_0 S  V_\mathrm{eff}} \frac{1}{(2\pi)^4 }\int \dif^2 q_\parallel | \vec{f}_{\mathrm{cav},x}(\vec{q}_\parallel)|^2+| \vec{f}_{\mathrm{cav},y}(\vec{q}_\parallel)|^2 \\
    & = \frac{e^2 \pi \omega_\mathrm{cav} l_B^2 N_L \nu }{2  \hbar \epsilon_0 S   V_\mathrm{eff}}\int \dif^2 r_\parallel | \vec{f}_{\mathrm{cav},x}(\vec{r}_\parallel)|^2 + | \vec{f}_{\mathrm{cav},y}(\vec{r}_\parallel)|^2.
\end{align}
In the first step, we used that the field mode is longitudinal and therefore parallel to $\vec{q}_\parallel$ and defined $\vec{f}_\mathrm{cav}(\vec{r}_\parallel) = \vec{f}_\mathrm{cav}(\vec{r}_\parallel, z=0)$. Using $N_L = g_D S / (2\pi l_B^2)$ and defining
\begin{align}
     V_z = \frac{V_\mathrm{eff}}{\int \dif^2 r_\parallel | \vec{f}_{\mathrm{cav},x}(\vec{r}_\parallel)|^2 + | \vec{f}_{\mathrm{cav},y}(\vec{r}_\parallel)|^2},
\end{align}
we obtain the light-matter coupling of the Hopfield model
\begin{align}\label{eq:cupling}
    g = \sqrt{\frac{ e^2\omega_\mathrm{cav}  g_D \nu}{4 \pi \hbar \epsilon_0   V_z}}.
\end{align}
We see that $g$ can be obtained from a simulation of the relevant part of the effective mode volume $V_z$. Before evaluating it for the resonator under consideration here, we compare our result to that obtained in Ref.~\cite{hagenmuller_cavity_2012}. Here, the coupling of graphene Landau levels to the lowest mode of a Fabry-Perot cavity with length $L_z$ was considered. In the Coulomb gauge, the following coupling constant was derived (adapting the notation): 
    \begin{align}\label{eq:coupling_cg}
    g_{cg} = \frac{\omega_\mathrm{cyc}}{\omega_\mathrm{cav}}\sqrt{\frac{ e^2 \omega_\mathrm{cav}^2  g_D \nu}{4 \pi^2 \hbar \epsilon_0   \sqrt{\epsilon} c }}.
\end{align}
Using that for a Fabry-Perot cavity $\omega_\mathrm{cav} = \pi c /(\sqrt{\epsilon}L_z)$ and that the Coulomb and multipolar coupling constants are related by $ g_{cg}  = (\omega_\mathrm{cyc}/\omega_\mathrm{cav}) g$, we find that our result in Eq.~\eqref{eq:coupling} agrees with the one derived in Ref.~\cite{hagenmuller_cavity_2012} in Eq.~\eqref{eq:coupling_cg} after identifying $V_z = \epsilon L_z$. 

If we approximate $\vec{f}_\mathrm{cav}(\vec{r}) =  \widetilde{f}_\mathrm{cav}(\vec{r}_\parallel) \zeta(z) \vec{e}_\parallel $, so that the field mode has only in-plane polarization and factorizes into $z$ and $\vec{r}_\parallel$ dependent parts, respectively, we get 
\begin{align}
     V_z = \frac{1}{\zeta(z=0)} \int \, \dif z \epsilon(z)| \zeta(z) |^2\approx \epsilon \delta z,
\end{align}
and
\begin{align} \label{eq:coupling}
    g = \sqrt{\frac{ e^2 \omega_\mathrm{cav}  g_D \nu}{4 \pi \hbar \epsilon_0   \epsilon \delta z}}.
\end{align}
This is a very crude approximation that, as we show below, is not applicable to the setup studied in this work, but shows the connection of the coupling derived here in Eq.~\eqref{eq:cupling} to the result obtained in Ref.~\cite{hagenmuller_cavity_2012}.

\subsection{Finite Element Simulations for Mode Volume calculation}

The parameter $V_z$ is crucial for determining the coupling strength $g$ between a subwavelength resonator and the graphene charge density. It relates the full mode volume to the relevant fraction of the field distribution of the mode at the position of the matter system. Simulating $V_z$ enables us to translate the light-matter coupling strength from a Fabry-Pérot treatment, wherein $\omega_{\text{cav}}$, and therefore $g$, are determined by the spatial extent of the cavity mode, to that of a subwavelength resonator employed here.

We have therefore performed 3D finite element simulations in COMSOL Multiphysics of the bare cavity. The resonator is modeled as a unit cell of a periodic array of resonators. The resonator itself has the same shape and dimensions as the fabricated resonator shown in the main manuscript. The gold resonator plane is modeled by a "Transition Boundary Condition" with a thickness of \SI{200}{\nano \meter} and Au material properties. The Si/SiO$_2$ substrate is modeled as bulk Si, on top of which the resonator rests. We further model the \SI{2}{\micro \meter} thick BCB layer on top of the resonator plane with a static permittivity of $\epsilon=2.4$, which is followed by another bulk Si layer, representing the Si lenses (without taking into account any focusing effect of the lenses themselves).  We include the conductive back-gate at a distance of d = \SI{190}{\nano \meter} below the resonator plane, according to the real sample geometry.  

We then extract the overall mode volume $V_{\text{eff}}$ by exporting the 3D electric field components, which we plot in the y-z plane (viewing along the resonator gap) at the resulting resonance frequency $f = $ \SI{1.34}{\tera \hertz} in Figure \ref{supp-fig:modevol} for varying back-gate conductivities. The back-gate is represented by the white dotted line, whereas the resonator is depicted with the black solid lines, with the capacitive resonator gap located in the center. We note here, that the back-gate conductivity (modeled with a simple Drude model formalism) has a considerable effect on the mode distribution and quality factor of the resonator. This stems from the fact that the carriers in the conductive surface constitute a loss channel, which only slightly alters the cavity resonance frequency (due to it being modeled as infinitely thin), but forces the mode to decrease in amplitude via metallic boundary conditions. We assume the back-gate (Cr/Pt) conductivity as a free parameter, since bulk conductivities are not representative for thin-film metals such as the one used in our sample \cite{fuchs_conductivity_1938}. The sharp drop-off of the electric field \SI{2}{\micro \meter} above the resonator arises from the BCB-Si interface.

\begin{Sfigure}
    \centering
    \includegraphics[width =  \linewidth]{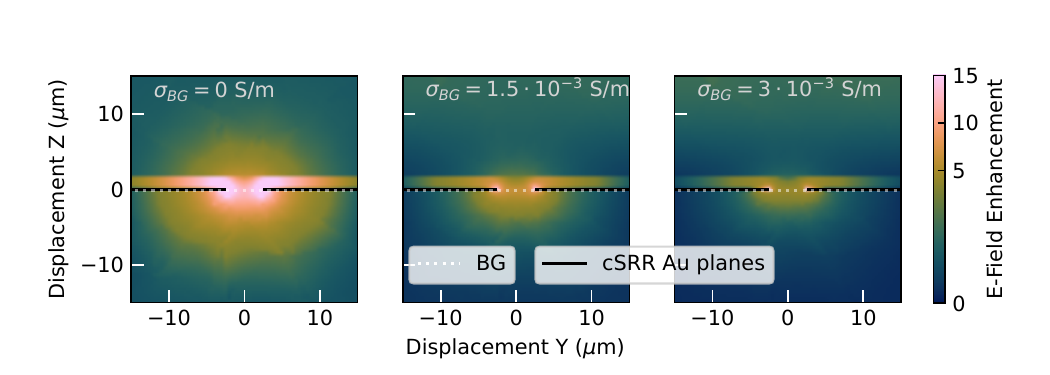}
    \caption{Simulated mode volumes of the subwavelength cavity at the cavity resonance frequency $f =$ \SI{1.34}{\tera \hertz} in the y-z plane (along the resonator gap) for various back-gate conductivities.}
    \label{supp-fig:modevol}
\end{Sfigure}

The calculated transmission spectra from the same simulations are overlaid with a measured transmission spectrum at CNP and the maximal measured magnetic field ($B=$ \SI{6}{T}). This measured spectrum represents the most cavity-like measurement of the sample, seeing as the coupling is minimal and the detuning of the cyclotron and cavity modes is maximized. Shown in Figure \ref{supp-fig:sparams}, with all spectra normalized to their respective maxima for comparison, the measured spectrum fits reasonably well with the simulated spectrum for a back-gate conductivity of $\sigma_\text{BG} = $ \SI{1.5}{\siemens \per \meter}. 

From the mode volume in 3D and the fraction of the in-plane electric field situated at the z-position of the graphene flake (not modeled in this simulation), we can quantify the parameter $V_z$ = \SI{0.16}{\milli \meter} which leads to the coupling strength curve plotted in the main manuscript, and represents the data without any fitting parameter.

\begin{Sfigure}
    \centering
    \includegraphics[width =  \linewidth]{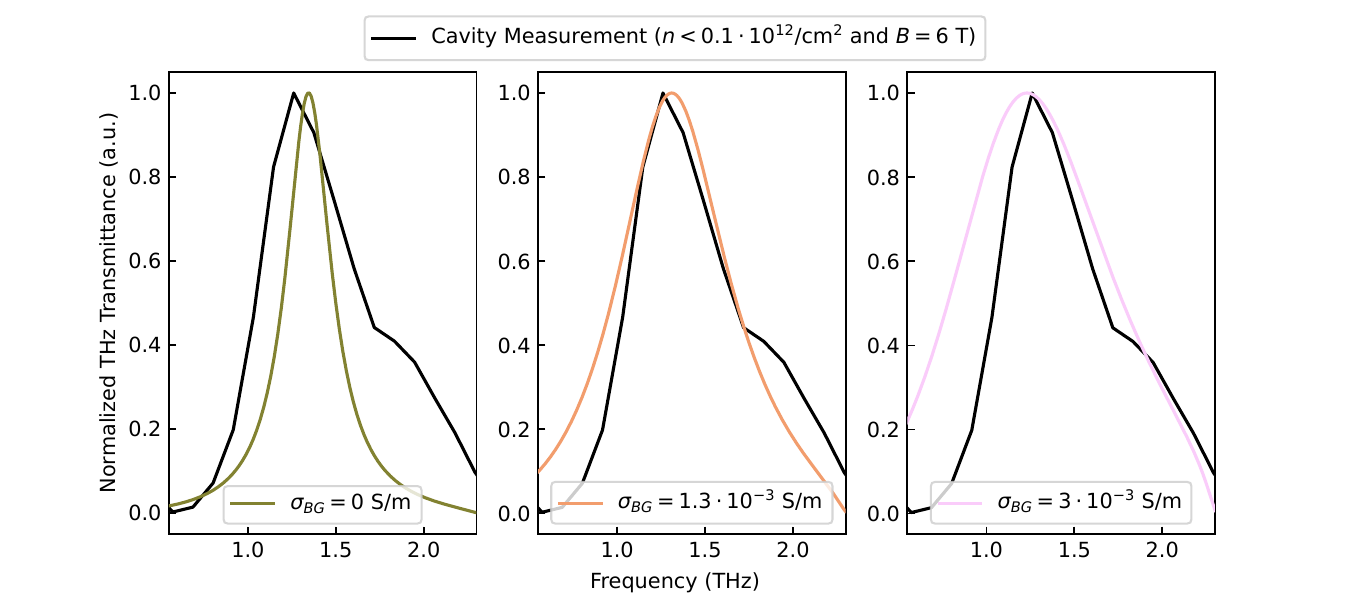}
    \caption{Simulated S-parameters (colorful lines) normalized and overlaid with a cavity-like measurement of the sample (black lines) for different back-gate conductivities.}
    \label{supp-fig:sparams}
\end{Sfigure}

\subsection{Interband transitions}

In Ref.~\cite{chirolli_drude_2012}, it was shown that for the coupling of a transverse cavity mode to the graphene Landau transitions treated in the minimal coupling scheme, coupling of the cavity to interband transitions cannot be neglected. If they are correctly incorporated, they dynamically lead to the $A^2$ term that is missing in the Hamiltonian otherwise, due to the linear dispersion in graphene. 

Here, we used a multipolar coupling scheme for which it was previously shown that a few-level truncation of the matter system works better than in the minimal coupling scheme \cite{de2018cavity,de2018breakdown}. This stems from the fact that in the former light couples to the position whereas in the latter it couples to the momentum operator of the charges. The different scaling of the matrix elements of the position and of the momentum operator with the corresponding transition frequency $\omega_\mathrm{ib}$ leads to the following relation between the coupling constants in the two coupling schemes: $ g_\mathrm{cg}  = (\omega_\mathrm{ib}/\omega_\mathrm{cav}) g_\mathrm{mc}$ (we denoted the multipolar coupling $g_\mathrm{mc}$ by just $g$ above). As we can see, in the multipolar coupling scheme the coupling of the cavity to high frequency off-resonant matter excitations with frequency $\omega_\mathrm{ib}$, such as interband transitions, is suppressed by a factor $\omega_\mathrm{cav}/ \omega_\mathrm{ib}$. This allows us to neglect interband transitions in the multipolar coupling scheme developed here. 

\newpage
\section{Superradiant phase transition of the Dicke model}
\subsection{General case}
In this section, we discuss the criticality of the Dicke model at zero and finite temperatures in a general setting.
The Dicke model in its original formulation \cite{dicke_coherence_1954,hepp_superradiant_1973} describes $N$ two-level emitters with transition frequency $\omega_0$ coherently coupled to a single quantized photonic mode at frequency $\omega_\text{cav}$. The Hamiltonian reads (we set $\hbar=1$ in this section for simplicity)
\begin{equation}
    H=\omega_\text{cav}a^\dagger a+\frac{1}{2}\sum_{i=1}^N\omega_0\sigma_{z,i}+\sum_{i=1}^N\lambda(a+a^\dagger)\sigma_{x,i},
\end{equation}
where $\sigma_{j,i}\text (j=x,z)$ denote the Pauli matrices for the $i$-th two-level system and $\lambda$ is the coupling strength of a single emitter. With the collective spin operator ($S=N/2$)
\begin{equation}
    \mathbf{S}=\frac{1}{2}\sum_{i=1}^N\boldsymbol{\sigma}_i,
\end{equation}
the Hamiltonian reads
\begin{equation}
    H=\omega_\text{cav}a^\dagger a+\omega_0S_z+2\lambda(a+a^\dagger)S_{x}.
\end{equation}
In the case of large $N\gg1$ and low excitation density $\langle S_z\rangle\ll N$, this Hamiltonian can be bosonized using a Holstein-Primakoff transformation
\begin{align}
    S_z&=b^\dagger b-\frac{N}{2}\\
    S_-&=\sqrt{N-b^\dagger b}b\approx\sqrt{N}b\\
    S_+&=b^\dagger\sqrt{N-b^\dagger b}\approx\sqrt{N}b^\dagger.
\end{align}
Using $S_x=\frac{1}{2}(S_++S_-$), the Dicke Hamiltonian then becomes 
\begin{equation}
    H\approx\omega_\text{cav}a^\dagger a+\omega_0b^\dagger b+ g (a+a^\dagger)(b+b^\dagger),
\end{equation}
where $g=\sqrt{N}\lambda$ is the collective light-matter coupling strength of $N$ emitters.
Here, we first focus on the case of constant $g$, which is more often discussed in the literature (cite some papers). In the next section, we discuss the case of Landau polaritons where $g\propto\omega_0$.

The polariton modes of the Dicke model are
\begin{align}
    \omega_\text{LP/UP} &= \frac{1}{2}\sqrt{\omega_\text{cav}^2 + \omega_0^2\mp \sqrt{\omega_\text{cav}^4 + 16g^2\omega_\text{cav}\omega_0 - 2\omega_\text{cav}^2\omega_0^2+ \omega_0^4} }.   
\end{align}

\begin{Sfigure}
    \centering
    \includegraphics[width=0.9\textwidth]{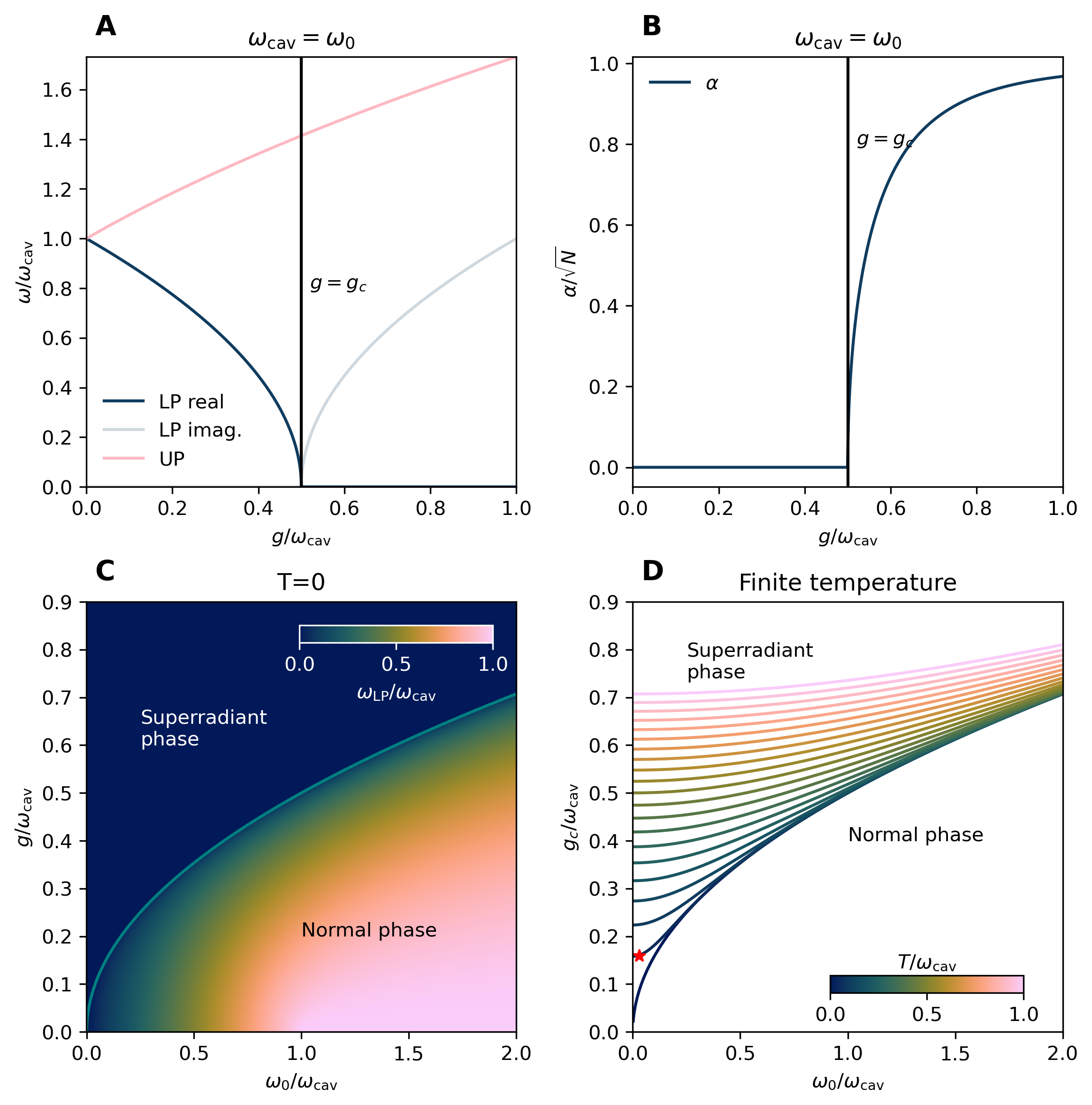}
    \caption{\textbf{Superradiant phase transition in the Dicke model.} \textbf{A} Polariton frequencies of the Dicke model at resonance. \textbf{B} Expectation value of the ground-state photon population. \textbf{C} Phase diagram of the Dicke Hamiltonian at zero temperature and fixed cavity frequency as a function of the matter frequency $\omega_0$ and coupling strength $g$. The color contrast corresponds to the LP frequency, which is zero in the superradiant phase and non-zero in the normal phase. The teal line marks the phase boundary. \textbf{D} Phase boundaries for the Dicke model at various finite temperatures $T\geq 0$. The red star marks the experimental temperature of 3 K.}
    \label{supp-fig:dicke}
\end{Sfigure}

The polariton branches are plotted at resonance ($\omega_\text{cav}=\omega_0$) as a function of the coupling strength in Fig. \ref{supp-fig:dicke}A.
Approaching the critical coupling
\begin{equation}
    g_c = \frac{\sqrt{\omega_\text{cav}\omega_0}}{2},
\end{equation}
the LP frequency softens and eventually vanishes at $g_c$. Beyond this point, the LP frequency becomes imaginary. The softening of the LP marks the onset of the superradiant phase, where the ground state develops a macroscopic photon population and the expectation value
\begin{equation}
    \alpha = \langle a\rangle
\end{equation}
becomes non-zero, as seen in Fig.\ref{supp-fig:dicke}B. This expectation value thus acts as the order parameter of the phase transition. Also note that since $E\sim a+a^\dagger$, the electric field $E$ develops non-zero expectation value in the superradiant phase. By looking at the LP (soft-mode) frequency, we can construct a phase diagram spanned by the tunable matter frequency $\omega_0$ and the coupling strength $g$ (see Fig. \ref{supp-fig:dicke}C). The normal and superradiant phases are separated by a square root-like phase boundary (teal line). Note that the superradiant phase can be entered either by changing the coupling strength or the detuning $\omega_0-\omega_\text{cav}$.

At finite temperatures $T>0$, we can employ mean-field theory to find the phase boundary \cite{kirton_introduction_2019}
\begin{equation}
    g_c(T) = \sqrt{\omega_\text{cav}\omega_0\coth{\frac{\omega_0}{2T}}},
\end{equation}
where we set the Boltzmann constant $k_B=1$ for convenience. Note that this function smoothly recovers the condition derived earlier as $T\to 0$. The corresponding phase boundaries at different temperatures are depicted in Fig. \ref{supp-fig:dicke}D. The experimental temperature of 3 K corresponds to the phase boundary marked by a red star and overlaps with the zero-temperature case for most of the phase space apart from very low frequencies.

\subsection{Landau polaritons}
\begin{Sfigure}
    \centering
    \includegraphics[width=0.9\textwidth]{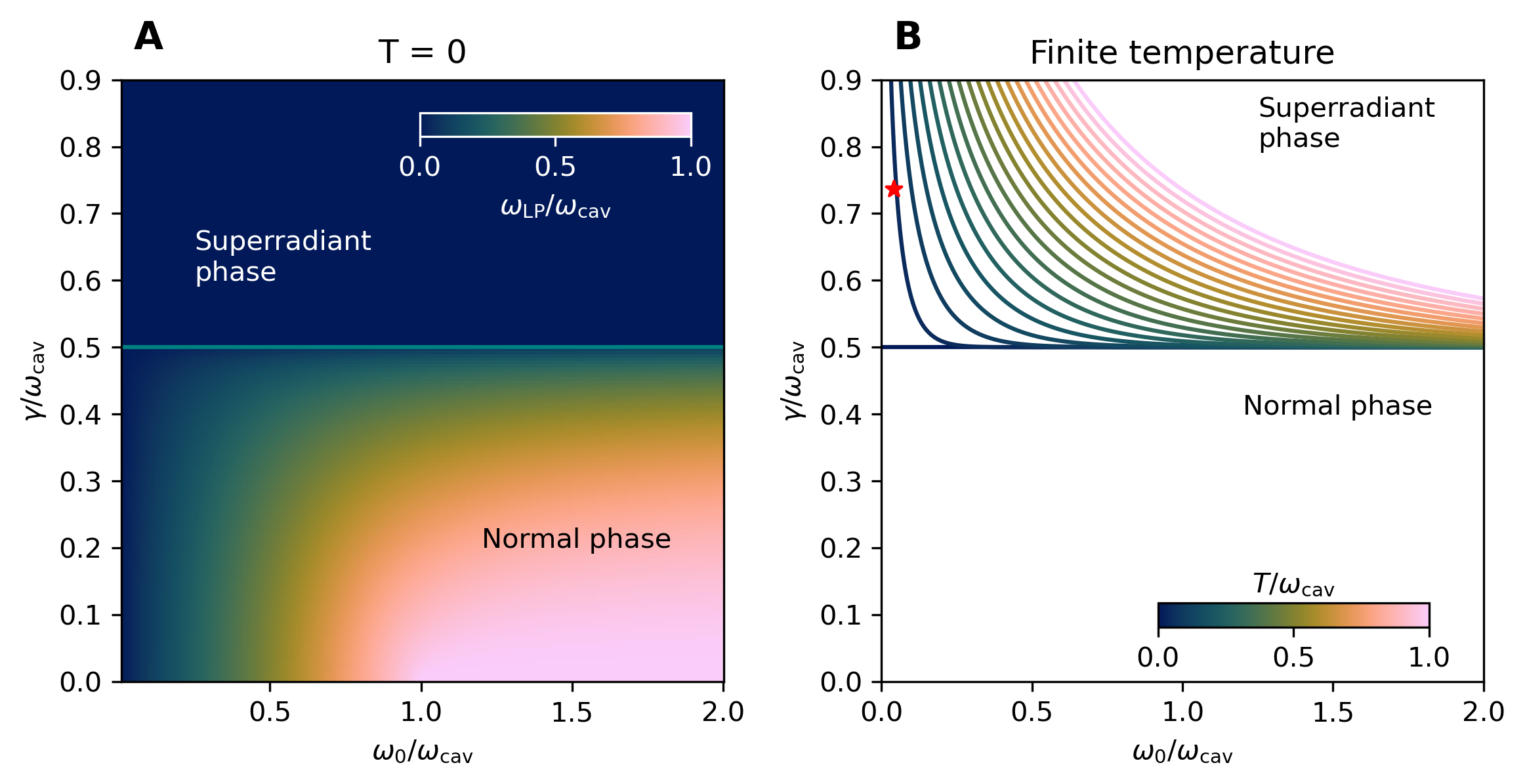}
    \label{supp-fig:dicke_landau}
    \caption{\textbf{Superradiant phase transition for Landau polaritons. } \textbf{A} Phase diagram of the Dicke Hamiltonian at zero temperature and fixed cavity frequency as a function of the matter frequency $\omega_0$ and coupling strength $g\propto \sqrt{\omega_0}$ (corresponding to Landau polaritons). The color contrast corresponds to the LP frequency, which is zero in the superradiant phase and non-zero in the normal phase. The teal line marks the phase boundary. \textbf{B} Phase boundaries for the Dicke model at various finite temperatures $T\geq 0$. The red star marks the experimental temperature of 3 K.}
\end{Sfigure}
Next, we focus on Landau polaritons relevant to our experimental work. In this case, the coupling strength $g$ depends as
\begin{equation}
    g=\gamma\sqrt{\frac{\omega_0}{\omega_\text{cav}}}
\end{equation}
on the cyclotron frequency, where $\gamma$ is the coupling at resonance $\omega_0=\omega_\text{cav}$. This changes the critical condition to 
\begin{equation}
    \gamma_c = \frac{{\omega_\text{cav}}}{2}.
\end{equation}
Importantly, the critical coupling is independent of the detuning between the cavity and the cyclotron frequencies. So, the hallmark of the superradiant phase transition in this system is the softening of the LP for all values of the magnetic field (equivalent to $\omega_0$) as a function of the bare coupling $\gamma$.

\bibliographystyle{naturemag}
\bibliography{bibliography}